\documentclass{aa}
\usepackage{graphicx}
\usepackage{natbib,twoopt}
\usepackage{nccmath}
\usepackage{nicefrac}
\usepackage{float}
\usepackage{bm}
\usepackage[utf8]{inputenc} 
\usepackage{multirow}
\usepackage{placeins}
\usepackage{textcomp}
\usepackage[breaklinks=True, colorlinks, citecolor=blue]{hyperref} 
\usepackage{chngcntr}
\usepackage{afterpage}
\counterwithout{figure}{section}
\counterwithout{table}{subsection}
\usepackage{placeins}
\usepackage{lipsum} 
\usepackage{stackengine}

\bibpunct{(}{)}{;}{a}{}{,}             
\makeatletter
  \newcommandtwoopt{\citeads}[3][][]{\href{http://adsabs.harvard.edu/abs/#3}%
    {\def\hyper@linkstart##1##2{}%
     \let\hyper@linkend\@empty\citealp[#1][#2]{#3}}}
  \newcommandtwoopt{\citepads}[3][][]{\href{http://adsabs.harvard.edu/abs/#3}%
    {\def\hyper@linkstart##1##2{}%
     \let\hyper@linkend\@empty\citep[#1][#2]{#3}}}
  \newcommandtwoopt{\citetads}[3][][]{\href{http://adsabs.harvard.edu/abs/#3}%
    {\def\hyper@linkstart##1##2{}%
     \let\hyper@linkend\@empty\citet[#1][#2]{#3}}}
  \newcommandtwoopt{\citeyearads}[3][][]%
    {\href{http://adsabs.harvard.edu/abs/#3}
    {\def\hyper@linkstart##1##2{}%
     \let\hyper@linkend\@empty\citeyear[#1][#2]{#3}}}
\makeatother

\bibpunct{(}{)}{;}{a}{}{,} 

\newfont{\gwpfont}{cmssq8 scaled 1000}

\begin{document}

\title{Towards mapping turbulence in the intra-cluster medium}
\subtitle{II. Measurement uncertainties in the estimation of structure functions}

\author{E.~Cucchetti\inst{\ref{inst1}} \and N.~Clerc\inst{\ref{inst1}} \and E.~Pointecouteau\inst{\ref{inst1}} \and P.~Peille\inst{\ref{inst2}}  \and F.~Pajot\inst{\ref{inst1}}}


\institute{
IRAP, Université de Toulouse, CNRS, CNES, UPS, (Toulouse), France 
\label{inst1}
\and CNES, 18 Avenue Edouard Belin 31400 Toulouse France \label{inst2}
}

\date{Received 2019}
\abstract{X-ray observations of the hot gas filling the intra-cluster medium (ICM) provide a wealth of information on the dynamics of clusters of galaxies. The global equilibrium of the ICM is believed to be ensured by \textcolor{black}{non-thermal and thermal pressure support sources, among which gas movements and the dissipation of energy through turbulent motions}. Accurate mapping of turbulence using X-ray emission lines is challenging due to the lack of spatially resolved spectroscopy. Only future instruments such as the X-ray Integral Field Unit (X-IFU) on \textsl{Athena} will have the spatial and spectral resolution to quantitatively investigate the ICM turbulence \textcolor{black}{over a broad range of spatial scales}. Powerful diagnostics for these studies are line shift and the line broadening maps, and the second-order structure function. When estimating these quantities, instruments will be limited by uncertainties of their measurements, and by the \textcolor{black}{sampling variance} (also known as cosmic variance) of the observation. Here, we extend the formalism started in our companion paper I to include the effect of statistical uncertainties of measurements in the estimation of these line diagnostics, in particular for structure functions.  We demonstrate that statistics contribute to the total variance through different terms, which depend on the geometry of the detector, the spatial binning and the nature of the turbulent field. These terms are particularly important when probing the small scales of the turbulence. An application of these equations is performed for the X-IFU, using synthetic turbulent velocity maps of a Coma-like cluster. Results are in excellent agreement with the formulas both for the structure function estimation ($\leq 3\%$) and its variance ($\leq 10\%$). The expressions derived here and in paper I are generic, and ensure an estimation of the total errors in any X-ray measurement of turbulent structure functions. They also open the way for optimisations in the upcoming instrumentation and in observational strategies.}
\keywords{Galaxies: intra-cluster medium - Instrumentation: \textit{Athena}/X-IFU - Lines: profiles - Methods: numerical - Techniques: imaging spectroscopy -  Turbulence - X-rays: galaxies: clusters}

\maketitle

\section{Introduction}

The X-ray emission of clusters of galaxies offers a phenomenal window to observe the thermodynamic and dynamic properties of the hot baryons composing the intra-cluster medium (ICM). The gas trapped in the dark matter potential of these structures holds an untouched fossil record of their formation, giving us a glimpse of the early days of the Universe \citep[see][for reviews]{Kravtsov2012Cluster,Planelles2016}. The first observations of the ICM showed smooth, spherical profiles, well described by $\beta$-models \citep{Cavaliere1978beta}, suggesting that the gas could be considered in (or close to) hydrostatic equilibrium. Several subsequent X-ray missions have since demonstrated that the ICM is far from homogeneous \citep{Fabian2006Perseus,Vikhlinin2006Temp,Leccardi2008Temps}. Dynamics induced by constant 3D accretion from the medium surrounding the cluster and by merger events throughout their lifetime are strengthened by the role of central active galactic nuclei (AGNs), whose jets, outflows, and bubbles, drive powerful mechanical and radiative motions, stirring the ICM at every spatial scale \citep{Fabian2012AGN,King2015Rev,Gaspari2017Feed,Morganti2017AGN}. Other effects present both at small (e.g. galaxy outflows) and large scales (e.g. sloshing, ram-stripping) also create heterogeneities in the gas emission, thereby severely questioning the assumption of hydrostaticity. 

Hints of systematic deviations from hydrostatic equilibrium up to a 10 to 20\% are indeed found in both state-of-the-art numerical simulations of the ICM and observational mass measurements \citep[see][for a review]{Pratt2019Rev}. \textcolor{black}{Other thermal and non-thermal pressure} support mechanisms are therefore called upon to compensate the cooling infall of the ICM towards the central parts of the clusters. The identification of the mechanisms responsible for such deviations are crucial to understand \textcolor{black}{the overall equilibrium of clusters} \citep[see][for a review]{Werner2019TurbRev}, and to have unbiased estimations of their mass, which is key to precision cosmology with clusters.

The dissipation of kinetic energy through either bulk or turbulent motions within clusters is a likely candidate \citep{Gaspari2018kine, Voit2018Turb}. In the classical view of the ICM, bulk motions are driven on a full-cluster scale by mechanisms such as mergers or ram-stripping, while turbulence indicates smaller scale motions. Turbulent energy is injected at hundreds of kpc scales and transported through a vortex cascade down to tens of kpc \citep{Donnert2018Scales}. \textcolor{black}{The gas motions induced by the turbulent cascade create a non-thermal pressure support mechanism \citep{Lau2009Mass}, while} the subsequent dissipation of the kinetic energy through collisions, small-scale viscosity, and eddies releases heat to the environment, counteracting part of the cooling flows in the cluster core \citep{Zhuravleva2014Fluct}. Scale-independent assumptions of the turbulent eddies naturally result in power-law forms of the turbulent power spectrum, characterised by an injection and dissipation scale of the energy, along with the characteristic slope of the spectrum \citep{Kolmogorov1941_1}. The determination of the energy injection scale provides information on the dominant energy transport mechanism at the cluster scale, involved for instance in the circulation of metals from the interstellar medium to the ICM \citep{Rebusco2006Transport}. Knowledge of the dissipation length provides instead insight on the viscosity of the ICM and collision mechanisms at small spatial scales \citep{Schekochihin2009Turb}. 

Yet, a direct observation of the ICM kinematics remains challenging. Random movements of gas particles related to turbulence create \textcolor{black}{line shifts, induce further broadening of the line, and can add skewness in the projection of the natural line profile along the line-of-sight}. The understanding and mapping of these effects therefore require the measurement of centroid, width, and shape of the emission lines with accuracies of a few tens of km/s over the full cluster scale for typical Fe $K_{\alpha}$ lines ($\sim 6.4$\,keV). 

Most of the current generation of X-ray instruments cannot provide spatially resolved high-resolution spectroscopy to this level of accuracy.  Insight on the ICM kinematics therefore relies on other physical quantities, such as the measurement of bulk motions using cold shock fronts \citep{Marevitch2007Bullet}, or the investigation of surface brightness, temperature, and density fluctuations in nearby clusters \citep{Churazov2003Fluct, Churazov2012Bright, Zhuravleva2015Fluct}. These results are nevertheless insufficient for a definitive understanding of the kinematic pressure support. A ground-breaking step forward was achieved with the soft X-ray spectrometer (SXS) onboard \textsl{Hitomi} \citep{Takahashi2018Hitomi}. Despite its short lifetime, the SXS mapped for the first time the turbulent velocity of the Perseus cluster, showing a quiescent ICM with velocities $\sim 200$\,km/s \citep{Hitomi2016Quiescent, Hitomi2017Gas}. New results are expected with the X-ray Imaging and Spectroscopy Mission (\textsl{XRISM}, \citealt{Ishisaki2018XARM}) and its instrument Resolve. However, a spatial mapping of turbulent velocity fields with accuracies of $\sim 10/20$\,km/s down to a few tens of kpc will require instruments such as the X-ray Integral Field Unit (X-IFU, \citealt{Barret2016XIFU,Barret2018XIFU}) on board the future X-ray observatory \textsl{Athena} \citep{Nandra2013Athena}. The X-IFU will provide an unprecedented 2.5\,eV spectral resolution below 7\,keV with a spatial accuracy of 5$^{\prime \prime}$ (over a $5^{\prime}$ equivalent field-of-view), enabling turbulence measurements through line broadening and deformations of the natural line profile \citep{Ettori2013Athena}.

With the advent of high-resolution X-ray spectroscopy, powerful line diagnostics can be used to investigate turbulent motions. These include the shift and broadening of a spectral line, and the computation of the structure function of the line-of-sight velocity field \citep{Inogamov2003SF}, related to the turbulent power spectrum of the ICM \citep{Zhuravleva2012SF}. Any measurement of these quantities will be limited by statistical uncertainties, linked to the observational set-up, and by the \textcolor{black}{sampling variance} (or `cosmic' variance) of the observation, \textcolor{black}{which refers to the intrinsic variations of a given diagnostic} related to the small number of observations of a random process. A theoretical understanding of these effects could provide a significant step forward in our knowledge of the usual line diagnostics used to study the turbulence of the ICM.

An analytical treatment of the cosmic variance is provided in our companion paper I (Clerc et al., sub., hereafter CL19), and used here. It allows the fast computation of estimates of the cosmic variance uncertainties, without using iterative Monte-Carlo techniques. In this paper, we extend this approach to include the contribution of statistics to the overall error estimation of the usual line diagnostics (line shift, line broadening and structure function) in the case of turbulence in the optically-thin emitting plasma of clusters of galaxies. Starting from formalism developed in \citet{Zhuravleva2012SF, ZuHone2016Hitomi}, we derive in the first part of this paper (Sect.~\ref{sec:estimation}) the errors associated to the previous line diagnostics, notably on the value and the variance of the estimated structure function. The formulas are generic, and remain valid for any level of statistical error obtained from measurements with an X-ray instrument. A specific application on the future X-IFU instrument is provided, on the basis of synthetic observations  (Sect.~\ref{sec:synth}) and the comparison of their post-processed outcomes with the prediction from our developed formalism  (Sect.~\ref{sec:cosmic}). The implications of these error formulas are then discussed, along with ways to estimate these contributions (Sect.~\ref{sec:discuss}). Throughout this paper, we assume a $\Lambda$-CDM cosmology, with $h=0.72$, $\Omega_m=0.24$ and $\Omega_{\Lambda}=0.76$. Bold, underlined letters $\underline{\vec{x}}$ indicate 3D vectors, bold letters $\vec{x}$ indicate 2D vectors. In a 3D space mapped by a $(x,y,z)$ orthonormal frame, $x$ is taken as the line-of-sight direction and $(y,z)=\vec{\theta}$ as the plane-of-sky coordinates, $\langle \cdot \rangle$ indicates the average operator, $||\cdot||_2$ the Euclidean norm and $\overline{X}$ an estimator of the quantity $X$. Other notations are consistent with paper I.

\section{Line diagnostics with finite statistics}
\label{sec:estimation}

\subsection{Line centroid and broadening}

Two tools to investigate the gas motions \textcolor{black}{projected} along the line-of-sight are the line shift, $\delta E$, and the line broadening, $\Sigma$. Line shift is defined as the difference between the energy of the line in the inertial frame of the observer, $E_0$, and the measured value. It can be related to either gas motions along the line-of-sight, or to the cosmological redshift $z$ of the source. By noting $E_{z}$ the energy of the line in the frame of the source, and $I_l(E)$ the line profile,\footnote{For instance a Gaussian or Voigt profile multiplied with line-of-sight emissivity.} $\delta E$ along a given line-of-sight $\vec{\theta}$ is defined as
\begin{equation}
\delta E(\vec{\theta}) = F^{-1}(\vec{\theta}) \int (E - E_{z}) I_l (E, \vec{\theta}) dE
\end{equation}
where $F(\vec{\theta})= \int I_l (E, \vec{\theta}) dE$ is the integrated flux of the line. Correspondingly, $\delta E$ can be expressed as a centroid velocity shift $C$ \textcolor{black}{of the projected line-of-sight component of the velocity field} as (with $c$ the speed of light):
\begin{equation}
C(\vec{\theta}) = \frac{\delta E(\vec{\theta})}{E_{z}} c
\end{equation}

The broadening of a line is the dispersion around its centroid value, and can be expressed similarly using
\begin{equation}
\Sigma^2(\vec{\theta}) = F^{-1}(\vec{\theta}) \int (E-\delta E- E_{z})^2 I_l (E,\vec{\theta}) dE
\end{equation}
Small turbulent motions create shifts in the corresponding line centroids of the gas particles. Integrated over along the line-of-sight, these result in a broadening of the observed line. The velocity broadening is thus defined by
\begin{equation}
\tilde{S}^2(\vec{\theta}) = \frac{\tilde{\Sigma}^2(\vec{\theta})}{E_{z}^2} c^2
\end{equation}
where $\tilde{\Sigma}$ is the broadening after subtraction of the instrument spectral resolution and other physical broadening effects (e.g. thermal broadening), assumed perfectly known here. The measurement of $\tilde{S}$ provides insight on the velocity distribution along the line-of-sight, making it a tool widely used to study turbulence \citep{Hitomi2017Gas}.

\subsection{The structure function}

Another line diagnostic tool for turbulence is the structure function. Its use in turbulence analysis originates from the early studies of turbulent motions in fluid dynamics \citep{Kolmogorov1941_2} before its extension to other branches of science (e.g. Earth sciences under the name of `variogram'), and later to astrophysics (\citealt{Miville1995Turb} in studies of the interstellar medium, \citealt{Roelens2017SF} in stellar variability, \citealt{Martinez2010SF} for galaxy clustering, or \citealt{Inogamov2003SF} in the case of ICM turbulence). The structure function appears when observing the dispersion $\sigma$ of \textcolor{black}{the line-of-sight component $v$ of the velocity field} over all the points in space $\underline{\vec{x}} \in \mathbb{R}^3$
\begin{equation}
\sigma^2 = \langle v^2(\underline{\vec{x}})  - \langle v(\underline{\vec{x}}) \rangle_x^2 \rangle_x = K_v(\underline{\vec{0}})
\end{equation}
which is a particular case of the auto-covariance function of the velocity field, $K_{v}$. \textcolor{black}{Under the assumption of an isotropic velocity field, the second-order structure function of the 3D field $v$, $\mathcal{SF}_2$, can be expressed exclusively as a function of spatial separation $s$ between two points in space, and is related to $K_v$ by}
\begin{equation}
\begin{split}
\mathcal{SF}_2 (s) &= 2 (K_v(\underline{\vec{0}}) - K_{v}(\underline{\vec{r}})) \\
&= \langle (v(\underline{\vec{x}} + \underline{\vec{r}}) - v(\underline{\vec{x}}))^2 \rangle_x
\end{split}
\end{equation}
where we average over all points $\underline{\vec{x}} \in \mathbb{R}^3$ separated by a distance $||\underline{\vec{r}}||_2=s$. The measurement of $\mathcal{SF}_2$ provides a view of the underlying turbulent velocity power spectrum through a `modified' second order moment of the velocity field. \textcolor{black}{Although the properties of a turbulent field are not fully characterised by its power spectrum, the properties of structure functions and their simple estimation} in fluid dynamics explains the success of this approach in all turbulence-related subjects. Notably, $\mathcal{SF}_2$ can be used to estimate the characteristic lengths of the turbulence \citep{Miniati2015Turb}. More generally, we can define the $n^{\rm th}$ moment of the structure function ($n \in \mathbb{N}$) as 
\begin{equation}
\mathcal{SF}_n (s) = \langle (v(\underline{\vec{x}} + \underline{\vec{r}}) - v(\underline{\vec{x}}))^n \rangle_x
\end{equation}
for a separation $||\underline{\vec{r}}||_2=s$. In the following sections, $\mathcal{SF}$ indicates the second-order structure function, and $\mathcal{D}$ the first-order structure function, or incremental function.

\subsection{Estimators and value: the influence of finite statistics}

The measurement of a velocity shift or a velocity broadening is related to a choice of the line-of-sight. Similarly, the definition of the structure function is related to a spatial average, such that an exact value can only be accessed either by averaging over a large number of spatial data points, or -- if we assume ergodicity -- by averaging over the same area for a large number of realisation of the turbulent velocity field. When observing astrophysical sources only a finite number of pointings and a limited exposure time are possible. Hence, it is essential \emph{to distinguish, for any line diagnostic, between the true value  and its estimation}. 

\subsubsection{Definitions and estimators of the structure functions}

In the rest of this study, we assume ergodicity and isotropy of the turbulence processes. For a given point in space $\underline{\vec{x}} \in \mathbb{R}^3$ with a speed $\underline{\vec{v}}$ in the referential of the observer, we only consider the \textcolor{black}{velocity component along the line-of-sight,} that is $\underline{\vec{v}}(\underline{\vec{x}}) \cdot \underline{\vec{e}}_x = v(\underline{\vec{x}})$, with no loss of generality due to isotropy. 

\textcolor{black}{In astrophysical observations, velocities can only be measured in the 2D space of the detector. Per pixel, the result will be the projection of the line-of-sight component of the velocity field modulated by emissivity effects}. We define the subset $\mathbf{S}_s\subset \mathbb{R}^4$, which contains all the doublets in the plane $(\vec{x}$, $\vec{y})$ separated by exactly $s$ in the sense of the Euclidean norm.  By convention, $\mathbf{S}_0 =  \varnothing$. We also define $\mathbf{\tilde{S}}_s$ as the `halved' subset, not counting for $\vec{x} \leftrightarrow \vec{y}$ permutations. The cardinal of $\mathbf{\tilde{S}}_s$ is noted $N_{p}(s)$ and represents the number of evaluations of the spatial average at a separation $s$. \textcolor{black}{Under these assumptions, the previous estimators can be transposed to their 2D projected equivalents, SF and D, using the centroid shift $C$:}
\begin{equation}
\textrm{SF}(s) = \langle (C(\vec{x}) - C(\vec{y}))^2 \rangle_{\mathbf{\tilde{S}}_s}
\end{equation}
\begin{equation}
\textrm{D}(s) = \langle C(\vec{x}) - C(\vec{y}) \rangle_{\mathbf{\tilde{S}}_s}
\end{equation}
where $\langle \cdot \rangle_{\mathbf{\tilde{S}}_s}$ is the average over the data points in $\mathbf{\tilde{S}}_s$. In practice, for any pixel (or centre of a region of pixels) in an observation, SF and D are computed using the following estimators:
\begin{equation}
\overline{\textrm{SF}}(s) = \frac{1}{N_{p}(s)} \sum_{({\vec{x}}, {\vec{y}}) \in \mathbf{\tilde{S}}_s} (\overline{C}({\vec{x}}) - \overline{C}({\vec{y}}))^2
\end{equation}
\begin{equation}
\overline{\textrm{D}}(s) = \frac{1}{N_{p}(s)} \sum_{({\vec{x}}, {\vec{y}}) \in \mathbf{\tilde{S}}_s} (\overline{C}({\vec{x}}) - \overline{C}({\vec{y}}))
\end{equation}
where  $\overline{C}$ is the estimator of the centroid shifts (i.e. actual measurement per pixel). In real data sets, only one, or a few realisations of these quantities will be computed. To determine whether these estimators are biased, one has to compute their expected value (in the statistical sense) and compare it to the real value. 

X-ray observations of the centroid shift and line broadening will be affected by sources of statistical and systematic errors, such that in every pixel or region $\vec{x}$, $\overline{C}({\vec{x}})=C({\vec{x}}) + \delta C({\vec{x}})_{\rm stat} + \delta C({\vec{x}})_{\rm syst}$ and $\overline{\tilde{S}}({\vec{x}})=\tilde{S}({\vec{x}}) + \delta \tilde{S}({\vec{x}})_{\rm stat} + \delta \tilde{S}({\vec{x}})_{\rm syst}$. The former is related to the exposure time of the observation, the latter to uncertainties in the calibration (energy scale and energy redistribution function) or in the fit. We assume no systematic error is present in the observations. The statistical error on each point is represented as a random variable, normally distributed and centred. We define $\sigma_{\textrm{stat}, C}$ and $\sigma_{\textrm{stat}, \tilde{S}}$ the standard deviation of the statistical error for $C$ and $\tilde{S}$ respectively (not necessarily equal). This distribution is considered spatially independent (i.e. valid on any subset of pixels). We further assume that the bulk motion is perfectly known, such that its contribution can be systematically subtracted from the measurements. Any turbulent velocity field is thus considered as centred, with a dispersion $\sigma_{\rm turb}$. 

\subsubsection{Expected value of the velocity shift and broadening}

The estimator of the centroid shift along a given line-of-sight is obtained directly from the measurements. The expected value of the velocity shift for a pixel (or region) $\vec{x}$ over multiple realisations of the same random process is simply (for a centred field):
\begin{equation}
\mathbb{E}[\overline{C}(\vec{x})] = 0
\end{equation}
The corresponding variance is (Appendix~\ref{subapp:line}):
\begin{equation}
\textrm{Var}[\overline{C}(\vec{x})] = \textrm{Var}[C(\vec{x})] + \sigma_{\textrm{stat}, C}^2
\end{equation}
where Var$[C$] is the \textcolor{black}{intrinsic variance of the centroid shift of the projected line-of-sight component of the velocity field} over different random observations, which is affected by emissivity (see CL19).

Similarly, the estimator of the line broadening is simply the measured broadening of the line. After subtraction of the instrumental effects and other physical effects, the estimator may be affected by the statistics of the measurements such that over multiple realisations of the velocity field (see Appendix~\ref{subapp:line}):

\begin{equation}
\begin{split}
\mathbb{E}[\overline{\tilde{S}^2}(\vec{x})] &= \mathbb{E}[\tilde{S}^2(\vec{x})] + \sigma_{\textrm{stat}, \tilde{S}}^2 \\
\textrm{Var}[\overline{\tilde{S}^2}(\vec{x})] &= \textrm{Var}[\tilde{S}^2(\vec{x})] +  4 \sigma_{\textrm{stat}, \tilde{S}}\mathbb{E}[\tilde{S}^2(\vec{x})] + 2\sigma_{\textrm{stat}, \tilde{S}}^4
\end{split}
\end{equation}
where $\mathbb{E}[\tilde{S}^2(\vec{x})]$ and $\textrm{Var}[\tilde{S}^2(\vec{x})]$ are respectively the expected value and the intrinsic \textcolor{black}{variance of the line broadening of the projected line-of-sight component of the velocity field} (see CL19). Statistics induce an additional broadening, which adds to the intrinsic variance through statistical terms and cross products.

\subsubsection{Expected value of the structure function}
\label{subsec:esf}

In the case of the structure function an ampler analytical approach is needed to quantify the effect of limited statistics in the measurements. For many observations of the same random turbulent process, the expected value of $\overline{\textrm{SF}}(s)$ is:
\begin{equation}
\mathbb{E}[\overline{\textrm{SF}}(s)] =   \sum_{({\vec{x}}, \vec{y}) \in \mathbf{\tilde{S}}_s}  \frac{\mathbb{E}[ (\overline{C}({\vec{x}})  - \overline{C}(\vec{y}))^2 ]}{N_{p}(s)}
\end{equation}

The development (see Appendix~\ref{app:computation}) yields the biased expected value of $\overline{\textrm{SF}}(s)$ shown by \citet{ZuHone2016Hitomi}:
\begin{equation}
\label{eq:esf}
\mathbb{E}[\overline{\textrm{SF}}(s)] = \textrm{SF}(s)+2 \sigma_{\textrm{stat}, C}^2
\end{equation}

Equation~\ref{eq:esf} -- valid throughout this paper -- shows that the measurement of the SF using a statistically inaccurate measurement of the turbulent velocity is systematically biased, regardless of the number of points used to derive the structure function. 

\subsubsection{Variance of the structure function}

It is important to determine whether the variance of the structure function is also affected by systematic biases. The accurate knowledge of the errors is crucial to understand the measurements and to distinguish between SF-related quantities (e.g. injection or dissipation scales). The same approach as in Sect.~\ref{subsec:esf} is thereby extended to the variance of the estimator:
\begin{equation}
\textrm{Var}[\overline{\textrm{SF}}(s)] =  \frac{1}{N_{p}(s)^2} \textrm{Var} \bigg [ \sum_{({\vec{x}}, \vec{y}) \in \mathbf{\tilde{S}}_s}  (\overline{C}({\vec{x}})  - \overline{C}(\vec{y}))^2 \bigg ]
\end{equation}

Under the previous assumptions, the variance of the estimator is given by (see Appendix~\ref{app:computation} for development):
\begin{equation}
\label{eq:varsf}
\begin{split}
\textrm{Var}[\overline{\textrm{SF}}(s)] &= \textrm{Var}[\textrm{SF}(s)] + \underbrace{4 \textrm{Var}[\textrm{D}(s)]\sigma_{\textrm{stat}, C}^2}_{\rm (1)} \\
&~~~ + \underbrace{\frac{4}{N_p(s)} \textrm{SF}(s)\sigma_{\textrm{stat}, C}^2}_{\rm (2)} + \underbrace{\frac{4 (N_{nei}(s)+1)}{N_{p}(s)} \sigma_{\textrm{stat}, C}^4}_{\rm (3)}
\end{split}
\end{equation}
where $N_{nei}(s)$ is the number of neighbours at a distance $s$ of any given point (see Appendix~\ref{subapp:nei} for a mathematical definition) and Var indicates the variance of the quantity over multiple observations of the same random process.

In the absence of statistical error (i.e. $\sigma_{\textrm{stat}, C}=0$), we recover the intrinsic variance of the structure function, which can be determined using the approach presented in CL19. With statistics, \textcolor{black}{we distinguish between three different terms}. Each can be interpreted as follows:
\begin{enumerate}[{(1)}]
\item \textbf{Velocity field fluctuation term:} The first term is related to the variance of the incremental function. This term provides a sense of the velocity fluctuations over the observational filed-of-view (FoV). If fluctuations are small (i.e. nearby pixels have similar velocities), the effect of a statistical error will be small. On the contrary, when pixel-to-pixel fluctuations are large, statistics will affect the computation of the estimator for a given observation of the turbulent velocity power spectrum. This term will therefore be small for dissipation scales larger than the pixel (or binned region) size, or large otherwise.
\item \textbf{Structure function fluctuation term:} This second term can be related to the uncertainty with which the structure function is computed when using turbulent velocities affected by statistical errors. Its value follows closely the shape of the `true' structure function (i.e. not positively biased, as in Sect.~\ref{subsec:esf}) and is therefore negligible at low spatial separations (where SF is small), but increases with $s$. This term becomes negligible when a large number of pairs is used to estimate the structure function.
\item \textbf{Statistical fluctuations term:} This term is the sheer contribution of the statistics to the overall variance of the structure function estimator. It appears from the covariance terms of the velocity field and is associated to the topology of the detector through the neighbour term $N_{nei}(s)$. Its contribution is most important at low spatial separations, where velocities are similar, and large statistical errors may introduce biases in the structure function estimation.
\end{enumerate}

\textcolor{black}{Alternatively, the first two terms can be written under a single term (1)+(2), related to the intrinsic nature of the line-of-sight component of the velocity field (both scale with $\sigma_{\textrm{stat},C}^2$). Formally, Var[D] is linked to SF (see Appendix~\ref{app:computation}) and both show similar properties, notably at large spatial scales. Whenever $N_p$ is large, terms (2) and (3), related to the number of regions used to evaluate SF, go to zero. The first term however, is intrinsically linked to the number of observations of the turbulent process, and remains non-zero, even for large $N_p$. It can be interpreted as a cross product between the cosmic variance of the field and the statistics, which can only be determined through multiple observations of the random process. For this reason, in the case of a Gaussian field, we made the choice to separate it from the sheer contribution of SF (2).} A verification of these formulas on simple test cases (e.g. constant velocity fields, Gaussian fields) yields excellent results.

\section{Generation of synthetic turbulent velocity fields}
\label{sec:synth}

Future micro-calorimeter instruments such as the X-IFU will provide the required spatially resolved high-resolution spectroscopy to measure line shifts or line broadening, thus setting constraints on the turbulent velocity fields of nearby clusters \citep{Roncarelli2018XIFU}. The typical measurable scales of turbulence span from the size of the FoV -- or the mapped area in case of multiple pointings -- to the size of the pixel. A study of the turbulent cascade over different spatial scales with X-ray instruments is therefore limited to nearby clusters, where kpc to Mpc scales are accessible with arcmin-like FoV and arcsec-like pixels (provided a sufficient angular resolution).

The assessment of turbulence in these objects, however, will be hindered by cosmic variance and further degraded by the limited statistics of the observations. We provide in the rest of this paper an application of the previous equations (Sect.~\ref{sec:estimation}) in the case of synthetic X-IFU observations to validate these formulas and demonstrate the capabilities of the instrument. The generation of the turbulent velocity fields is inspired from \citet{ZuHone2016Hitomi} (hereby Z16). Simulations are based on the official E2E simulator SIXTE \citep{Wilms2014SIXTE}, and performed similarly to \citet{Cucchetti2018Abun} (hereby C18).

\subsection{Emission profile and turbulent power spectrum}

Forecasted targets to investigate the ICM turbulence with the X-IFU are local and massive clusters, such as the Perseus and Coma clusters. Hereafter, we consider a Coma-like cluster, as in Z16, assuming an emission profile described by a $\beta$-model:
\begin{equation}
n_e(\underline{\vec{r}})=n_{e,0} \left(1+\left(\frac{||\underline{\vec{r}}||_2}{r_c} \right)^{2} \right)^{-\frac{3 \beta}{2}}
\end{equation}
where $n_{e,0}$ is the electron density at the core and $r_c$ the `core' radius of the cluster. In the rest of this paper we assume $n_{e,0} = 3 \times 10^{-3}\,$cm$^{-3}$, $r_c = 400\,$kpc, $\beta=2/3$, and $n_e=1.2\,n_H$. We consider observations of the core of the cluster, where emissivity varies but slightly ($\sim 2\%$) over an X-IFU FoV (i.e. $5^{\prime}$ in equivalent diameter). At the redshift of Coma ($z_0=0.023$), 1\,kpc corresponds to $2.21^{\prime \prime}$ on the sky. For simplicity, we assume an isothermal cluster at $k_B T=7$\,keV, with a constant metallicity $Z=0.7\,Z_{\odot}$ (see \citealt{Ettori2015evol}, abundances are given with respect to \citealt{Anders1989Solar}). The $x$ direction remains the line-of-sight of the observations and we chose the centre of the cluster as the origin of a $(x,y,z)$ orthonormal frame. For a point $\underline{\vec{r}}$, its 3D wave-vector is $\underline{\vec{k}}=(k_x, k_y, k_z)$, with $(k_y, k_z) = \vec{\xi}$.

\begin{figure*}[!t]
\centering 
\includegraphics[width=0.49\textwidth]{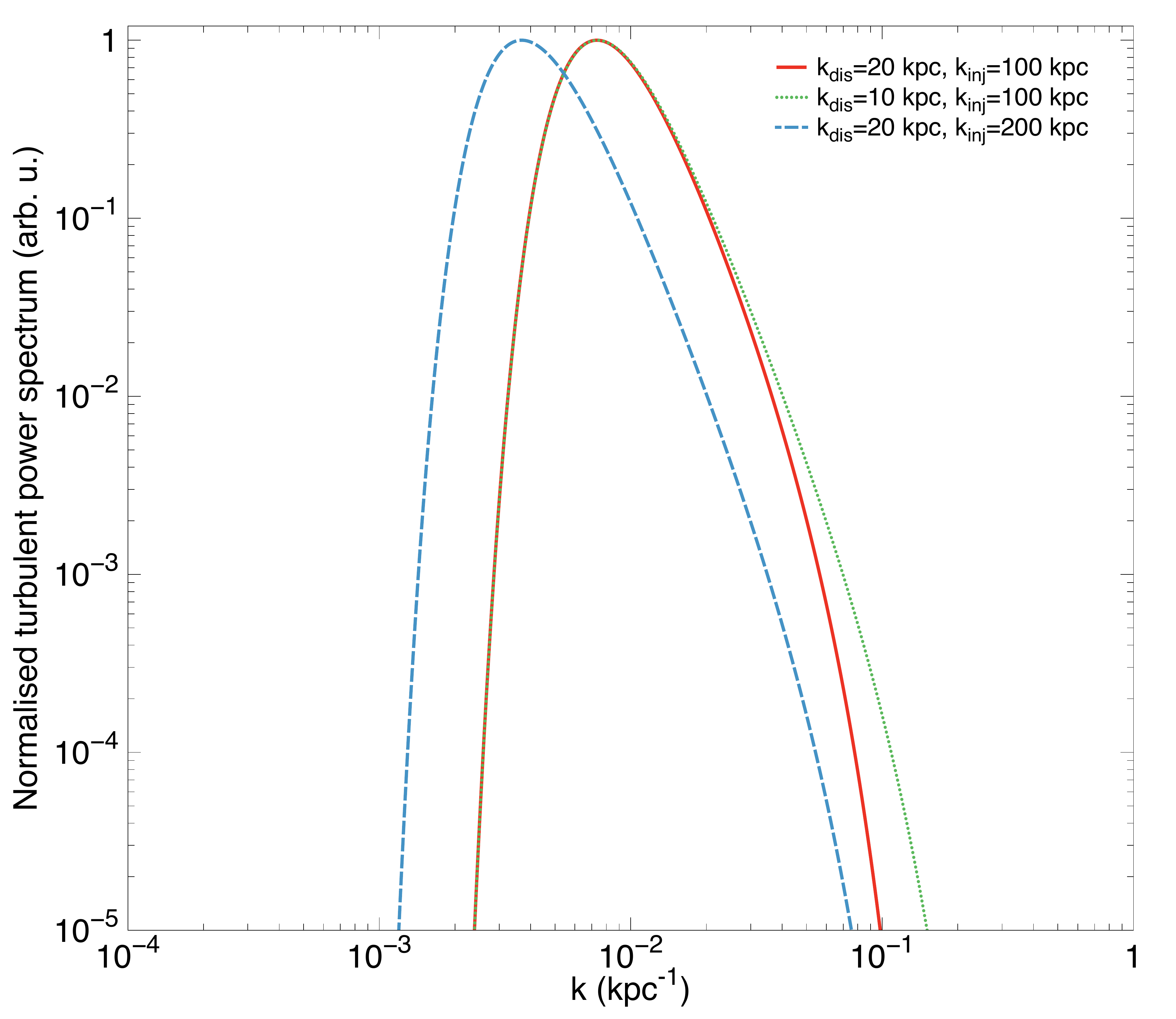}
\includegraphics[width=0.49\textwidth]{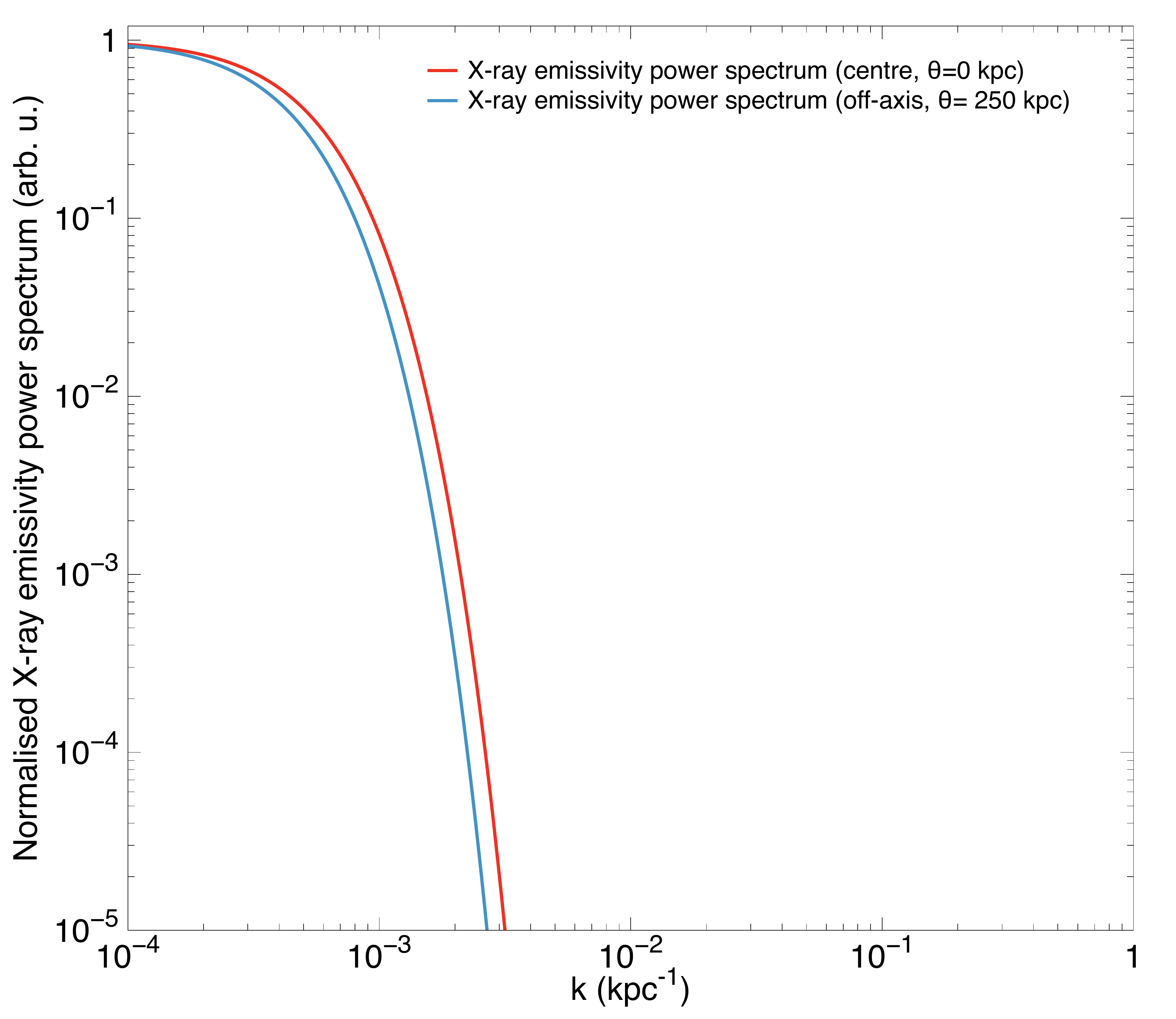}
\caption{Example of power spectra used in the simulations. (Left) Turbulent power spectra used in the simulation for different injections and dissipation scales. (Right) Normalised spectrum of X-ray emissivity of the Coma model at the centre used to compute $C_n$ ($||\vec{\theta}||_2=0$\,kpc, red) and off-axis ($||\vec{\theta}||_2 = 250$\,kpc, blue).}
\label{fig:pow}
\end{figure*}

Each gas particle in the ICM is simulated with a velocity $v(\underline{\vec{r}})$ along the line-of-sight. \textcolor{black}{A full description of turbulence requires hydrodynamical treatments \citep{Gaspari2013Turb, Gaspari2014Turb}. We simplify here this approach (also for computational reasons) by assuming that turbulence follows an isotropic Kolmogorov 3D power spectrum} 
\begin{equation}
P_{3D}(\underline{\vec{k}}) = ||\tilde{v}(\underline{\vec{k}})||_2^2 = C_{n} k^{\alpha} e^{-(k/k_{\rm dis})^2} e^{-(k_{\rm inj}/k)^2} 
\end{equation}
where $\tilde{v}$ is the 3D Fourier transform of the velocity field along line-of-sight, $C_n$ is a normalisation factor of the power spectrum and $k=||\underline{\vec{k}}||_2$. We note $k_{\rm dis}$, $k_{\rm inj}$ the dissipation and injection scale respectively, and $\alpha$ the turbulent power-slope (Figure~\ref{fig:pow} -- Left).

\subsection{Normalisation of the power spectrum}

Given the cluster emission profile, the velocity measured by the instrument will be convolved with the power spectrum of the cluster emission. As shown in Z16, if we note $\epsilon \propto n_e n_H \Lambda(T, Z)$ the X-ray volume emissivity, \textcolor{black}{the emission-measure weighted projection of the line-of-sight component of the velocity field along a given line-of-sight $\theta$ for a Gaussian or Voigt line is simply}
\begin{align}
C(\vec{\theta}) = \frac{\int v(\underline{\vec{r}}) \epsilon(\underline{\vec{r}}) dx}{\int \epsilon(\underline{\vec{\textbf{r}}}) dx} 
\end{align}
As the emission is isothermal and isometallic, $\epsilon$ only depends on the squared density of the cluster. By calling 
\begin{equation}
\label{eq:emi}
\rho(\underline{\vec{r}}) =  \frac{\epsilon(\underline{\vec{r}})}{\int \epsilon(\underline{\vec{r}})dx}
\end{equation}
the normalised X-ray emissivity, the velocity dispersion $\tilde{S}$ along a specific line-of-sight is given by
\begin{equation}
\tilde{S}^2(\vec{\theta}) = \int v(\underline{\vec{r}})^2 \rho(\underline{\vec{r}})dx - \left( \int v(\underline{\vec{r}}) \rho(\underline{\vec{r}}) dx \right)^2
\end{equation}

The expected value of $\tilde{S}^2$ along $\vec{\theta}$ is related to the turbulent power spectrum \citep[][CL19]{Zhuravleva2012SF} by:
\begin{equation}
\label{eq:Cn}
\mathbb{E}[\tilde{S}^2(\vec{\theta})] = \int P_{3D} (k_x, k_y, k_z) [1 - P_{\rho}^{~\vec{\theta}}(k_x)] dk_x dk_y dk_z
\end{equation}
where $P_{\rho}^{~\vec{\theta}}$ is the 1D power spectrum of the normalised X-ray emissivity $\rho$ (Equation~\ref{eq:emi}) for a fixed $\vec{\theta}$ (\textcolor{black}{depends on the selected line-of-sight}, see Figure~\ref{fig:pow} -- Left). As in Z16,  the normalisation of the 3D power spectrum is chosen to satisfy $\mathbb{E}[\tilde{S}^2(\vec{0})] =  (\mathcal{M} c_{\rm sound})^2$, where $\mathbb{E}[\tilde{S}^2(\vec{0})]$ is the \textcolor{black}{expected velocity broadening of the line-of-sight component of the velocity field at the centre of the cluster}, $\mathcal{M}$ the Mach number and $c_{\rm sound}$ the sound celerity in the ICM. For our simulations, we use $\mathcal{M} = 0.3$ and $c_{\rm sound}= 1460\,$km/s (Z16). 

The normalisation $C_n$ can be computed through numerical integration of Equation~\ref{eq:Cn} for $\vec{\theta}=\vec{0}$. In our study, the computation is performed using a uniform $8192 \times 8192 \times 8192$ grid of a length $k_{\ell} = 0.50$\,kpc$^{-1}$, which corresponds to the spatial scale $\ell$ of pixel. To accurately compute the small scales in the centre of the cluster (notably for $P_{\rho}^{~\vec{\theta}}$), we take $k_{\rm min}=(50r_c)^{-1}$\,kpc$^{-1}$ (Figure~\ref{fig:pow} -- Right). This approach yields excellent results with respect to the purely analytical approach (possible with a $\beta$-model) with accuracies better than $0.1\%$.

\subsection{Generation of the turbulent velocity field}

One realisation of the turbulent velocity field is generated under the previous assumptions using the 3D turbulent power spectrum.  We operate with a uniform area on the sky of $8.5^{\prime} \times 8.5^{\prime}$ (i.e. 230\,kpc $\times$ 230\,kpc) and 1.84\,Mpc along the line-of-sight. The $\vec{\theta}$ plane size is chosen to include more than 1.5 times the X-IFU FoV to avoid finite box-size effects of the simulation, while the grid is extended over the line-of-sight (8 times larger) to account more accurately for projection effects and ensure a smoother cut-off of the emissivity at the edges of the grid. For each run, we take a $2048 \times 256 \times 256$ mesh, with a step size of $\ell/2 = 2.14^{\prime \prime}$ (0.97\,kpc), which corresponds to the half-width of the X-IFU pixel to avoid aliasing (Shannon criterion), and offers a good computational speed compromise. 

Each grid point is given a turbulent velocity in Fourier space $\tilde{v} (\underline{\vec{k}}) = |V_k| e^{i \psi}$, with $|V_k|$ the modulus, and $\psi$ the phase, assumed without spatial correlation. As in Z16, we use a Rayleigh distribution of parameter $\Sigma_{V_k} = P_{3D}(\underline{\vec{k}})/2$ for the modulus, and a spatially independent uniform distribution of the phase $\psi$. The corresponding probability distribution function being
\begin{equation}
\mathcal{P}(V_k, \psi) dV_k d\psi = \frac{V_k}{\Sigma_{V_k} ^2} e^{-\frac{V_k^2}{2\Sigma_{V_k} ^2}} dV_k ~ \frac{d\psi}{2\pi}
\end{equation}

\begin{figure*}[!t]
\centering
\includegraphics[width=0.49\textwidth]{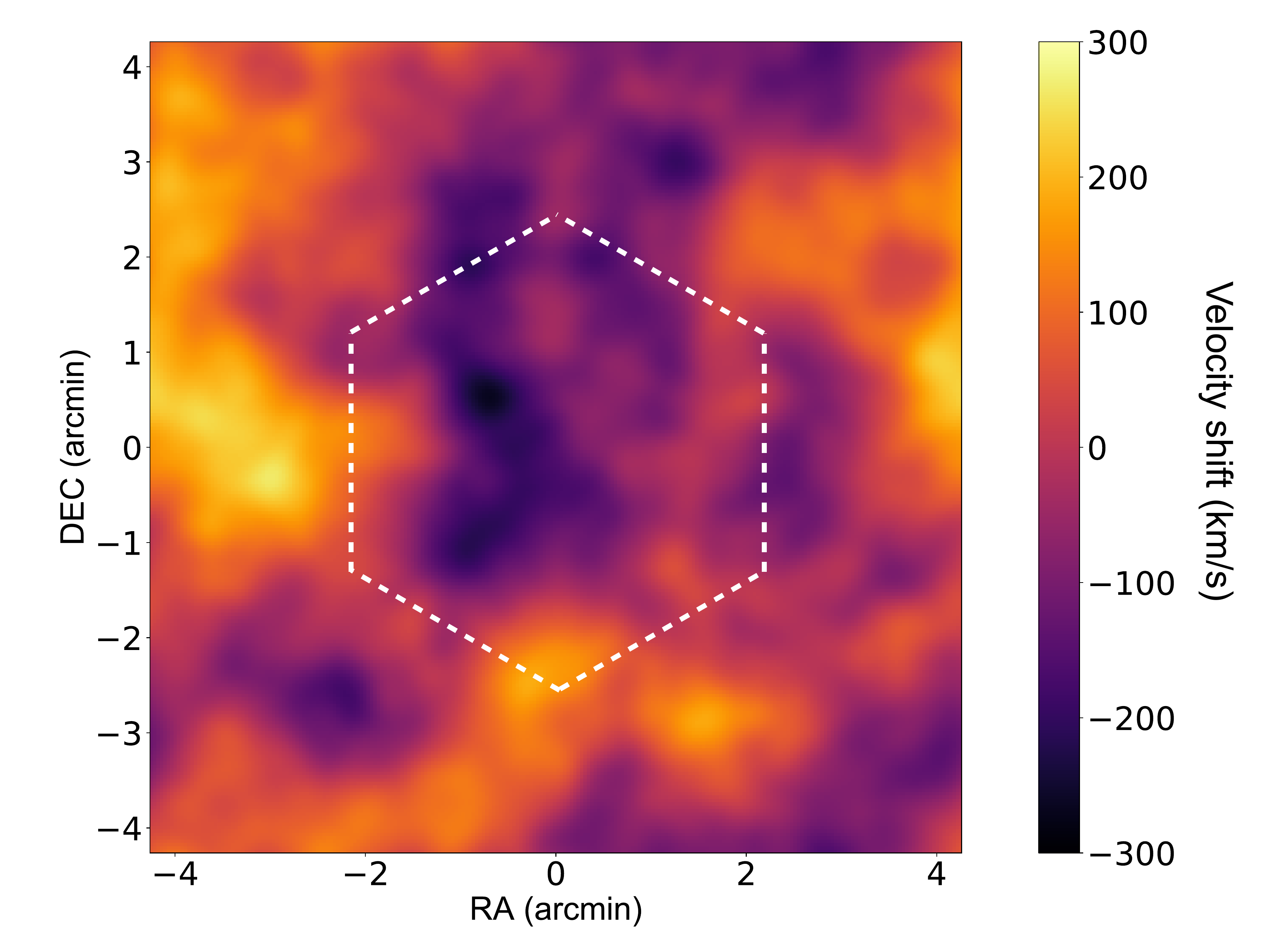}
\includegraphics[width=0.49\textwidth]{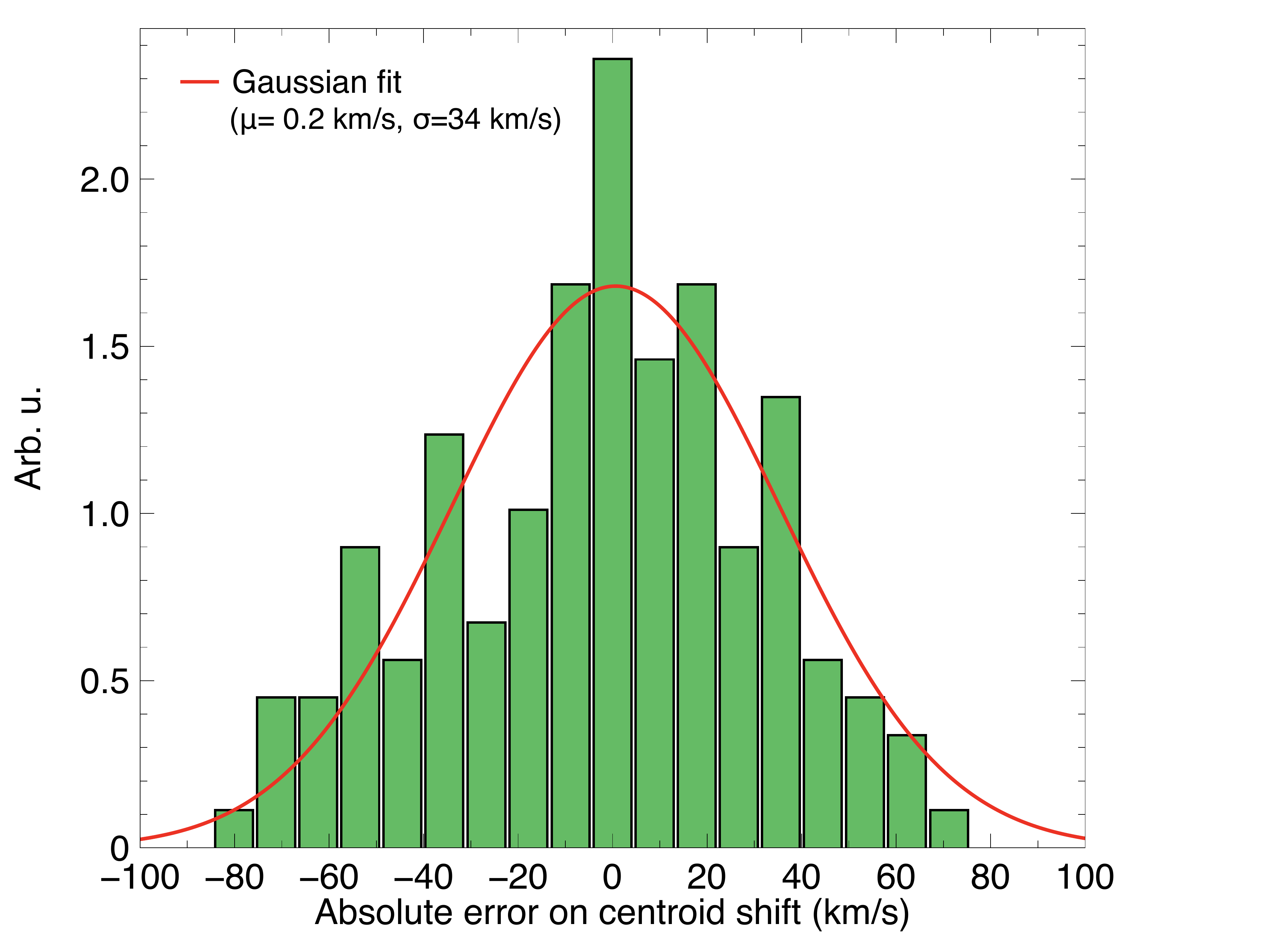}
\includegraphics[width=0.49\textwidth]{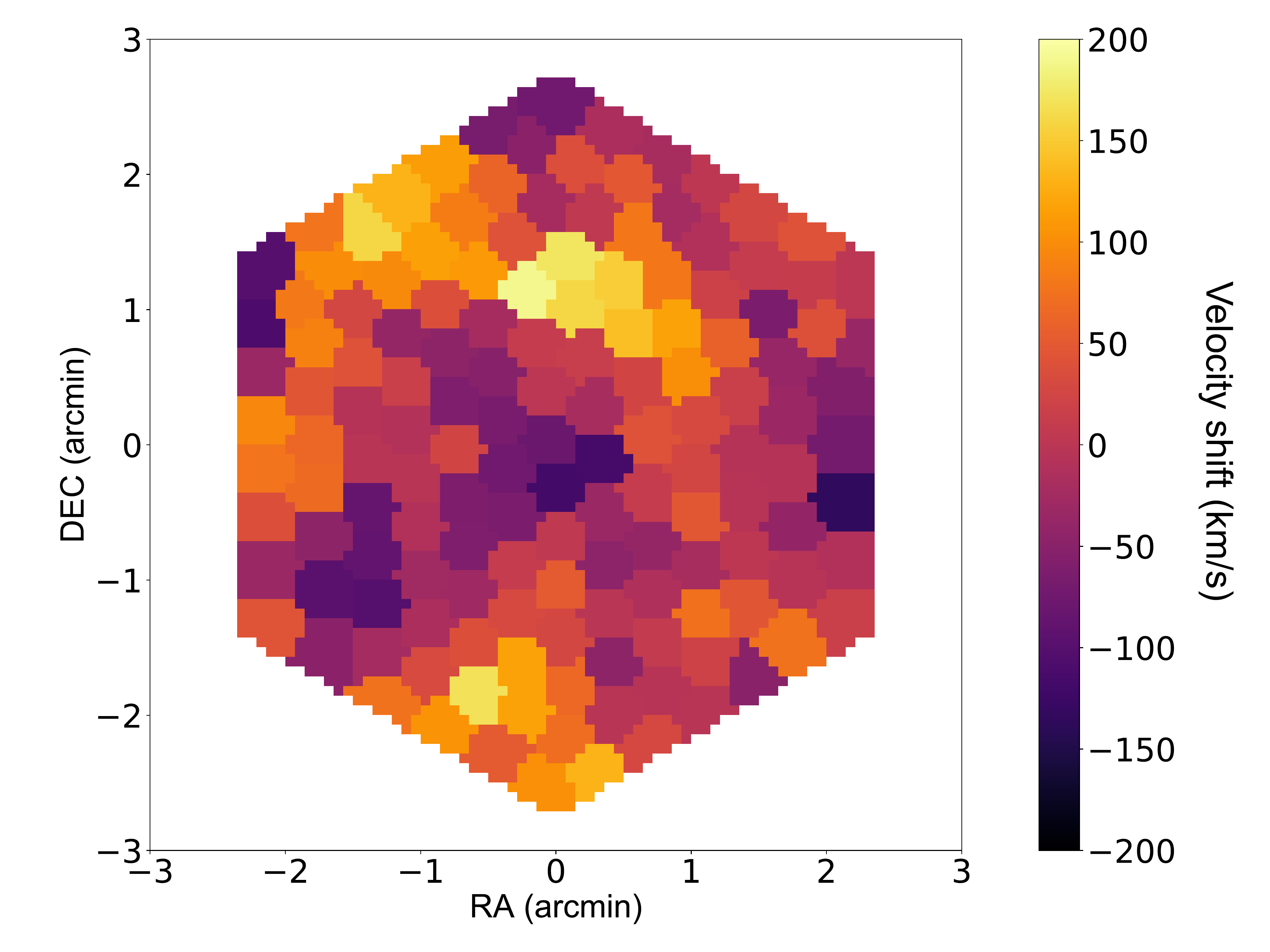}
\includegraphics[width=0.49\textwidth]{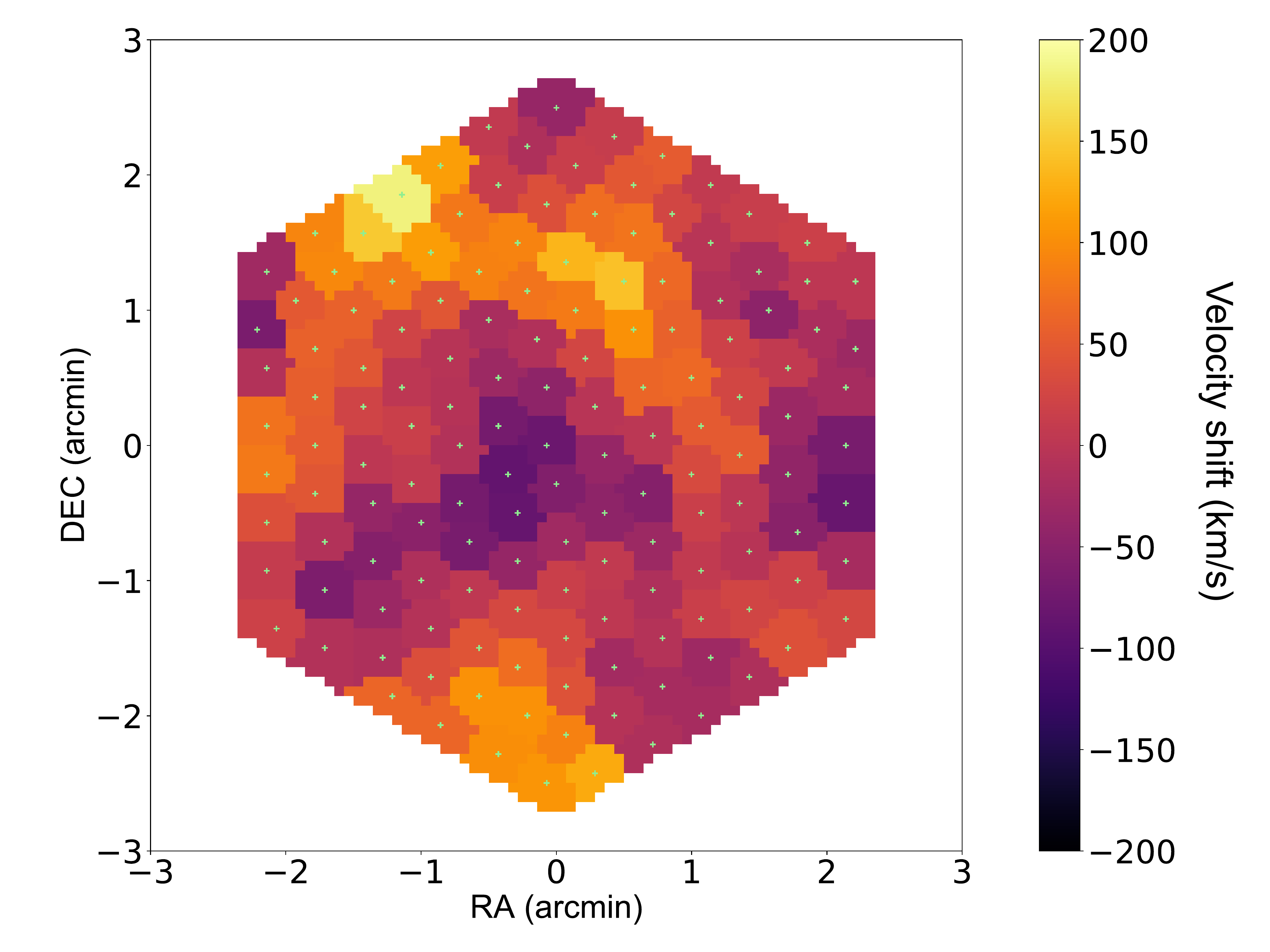}
\caption{Example of simulated velocity fields. (Top left) \textcolor{black}{Example of an emission-measure weighted projection of the simulated line-of-sight component of the turbulent velocity field.} In this case $k_{\rm inj} = 1/150$\,kpc$^{-1}$ and $k_{\rm dis} = 1/20$\,kpc$^{-1}$. The shape and the extent of the X-IFU FoV is shown as a white dashed line (Top right) Absolute error distribution between the recovered line-of-sight velocity in one of the simulations and the corresponding input emission-measure-weighted velocity. The statistical error follows a centred Gaussian distribution where the Gaussian best fit (in red) is found for $\mu_{\rm stat}=0.2$\,km/s and $\sigma_{\textrm{stat}, C} = 34$\,km/s. (Bottom left) Example of a synthetic X-IFU observation of bulk motion for $k_{\rm inj} = 1/200$\,kpc$^{-1}$ and $k_{\rm dis} = 1/10$\,kpc$^{-1}$ and (Bottom right) corresponding emission-measure weighted input map (binned). The small green crosses indicate the centres of the Voronoï regions. }
\label{fig:turbmap}
\end{figure*}

Without loss of generality, nor influence on the power spectrum, phases are computed exclusively on the lower triangular matrix of the velocity, and transposed to the upper triangular matrix to obtain a Hermitian velocity grid. Once computed in Fourier space, the velocity grid is transposed into real space \textcolor{black}{ to obtain the line-of-sight component of the velocity} $v(\underline{\vec{r}})$ in each point in the grid. Given the large arrays the inverse Fourier transform is performed through \texttt{2DECOMP\&FFT}\footnote{\tiny{\url{http://www.2decomp.org}}} \citep{Li2010Decomp}, which is memory-optimised for large matrices. \textcolor{black}{An example of the emission-measure weighted projection of a simulated velocity grid $v(\underline{\vec{r}})$} is provided Figure~\ref{fig:turbmap} (Top left).

\subsection{Simulation set-up}

The generated velocity grids can be used to validate the previous formulas in the case of X-IFU observations. They were therefore used as 
inputs to perform E2E simulations of our toy model Coma cluster.

\subsubsection{Particle emission model}

Similarly to C18, each gas particle is associated with a grid point and an element of volume, and assumed to emit isotropically. Since particle volumes are four times smaller than the X-IFU pixel area, and given the \textsl{Athena} telescope required angular resolution of 5$^{\prime \prime}$ half-equivalent width, we consider them as point-like individual sources on the sky. The emission spectrum for each particle is assumed to follow an unabsorbed thermally broadened plasma emission, modelled through XSPEC using \texttt{wabs*apec} with a constant temperature of 7\,keV and a metallicity of 0.7\,$Z{\odot}$. As in C18, the \texttt{wabs} absorption model \citep{MorrisonWabs} is preferred for computational efficiency. A column density value of $0.03 \times 10^{22}$\,cm$^{-2}$ is used, and represents a typical high Galactic latitude value of the column density seen over the sky \citep{Kalberla05Abs}. 

\textcolor{black}{The turbulent motions of the gas are included by converting the line-of-sight component of the velocity field for each grid point into an additional redshift}, which shifts the line by Doppler effect. No excess broadening is considered locally due for instance to microscopic turbulent motions. The corresponding \textcolor{black}{total} redshift $z_i$ of the $i^{\rm th}$ cell is computed using the classical redshift composition: 

\begin{align}
z_{i} = z_0 + z_{v,i} + z_0z_{v,i}
\end{align}
where $z_{v,i}$ is induced by the velocity of the cell,

\begin{align}
z_{v,i} = \sqrt{\frac{c+v_{i}}{c-v_{i}}}-1
\end{align}

Finally, the normalisation $\mathcal{N}_i$ of each emission spectra is provided for each cell through the \texttt{apec} normalisation 

\begin{align}
\mathcal{N}_i=\frac{10^{-14}}{4\pi[D_{\text{A}}(1+z_{x, i})]^2} n_{\text{e}} n_{\text{H}} V_{\rm cell}
\end{align}
where $D_A$ is the angular distance to the Coma cluster and $V_{\rm cell}$ the volume of each cell (constant for our uniform grid).

The photon generation and the simulation follow the same process as in C18, with the same \textcolor{black}{configuration of the} X-IFU instrument through SIXTE (\texttt{xifupipeline}). Since the main objective here is the estimation of turbulent velocities and the velocity power spectrum, no background nor cross-talk are included in the simulation to avoid introducing additional instrumental systematics. However, as shown in C18, an accurate knowledge of the background components should not bias the following results (especially for redshift measurements). Further, since the observations are focused on the centre of a bright Coma-like cluster, the astrophysical and instrumental background levels are expected to be sub-dominant with respect to the cluster emission. A typical exposure time of 100\,ks is considered for any of the following synthetic observations.
\begin{figure*} [!t]
\centering
\includegraphics[width=0.49\textwidth]{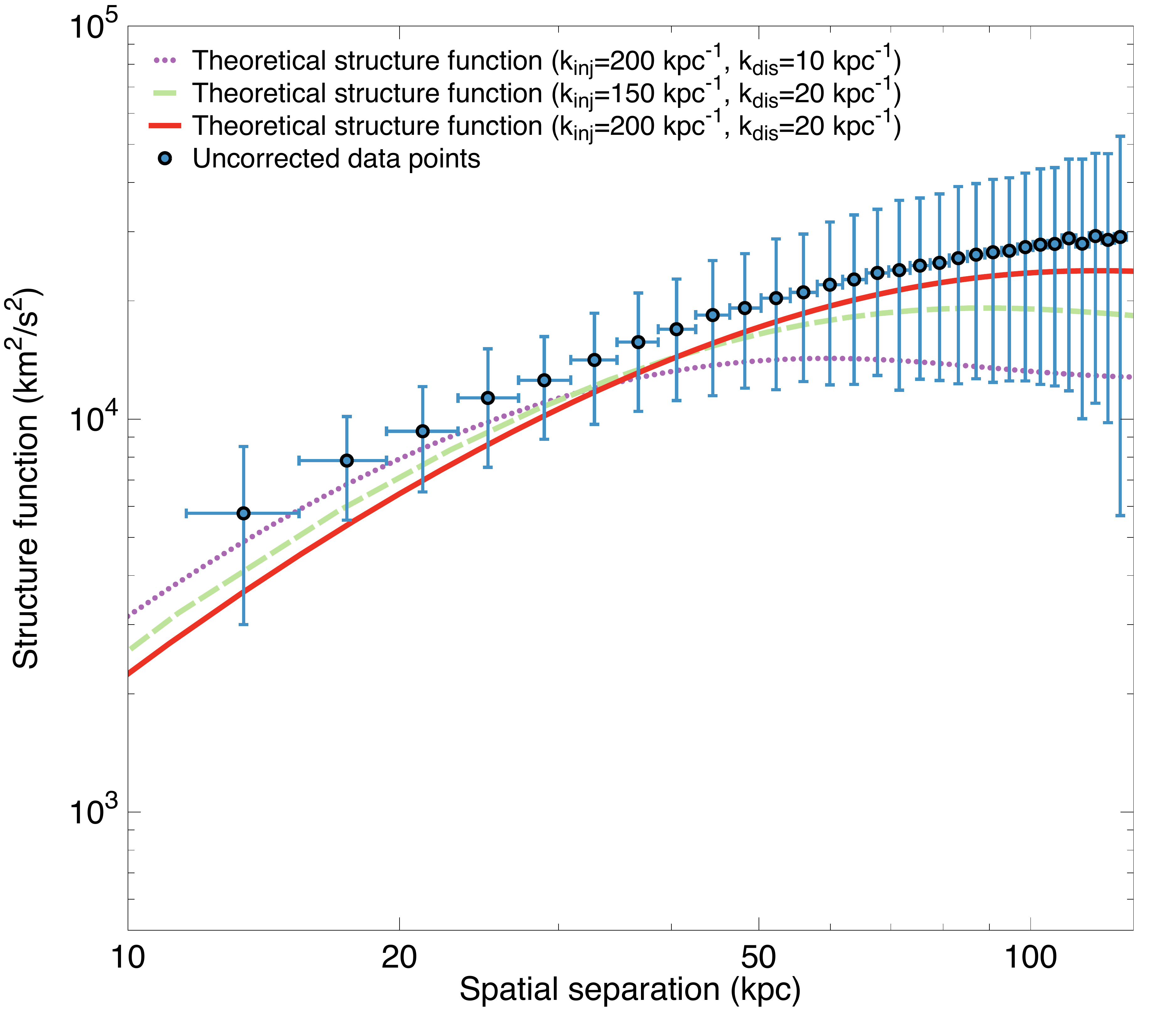}
\includegraphics[width=0.49\textwidth]{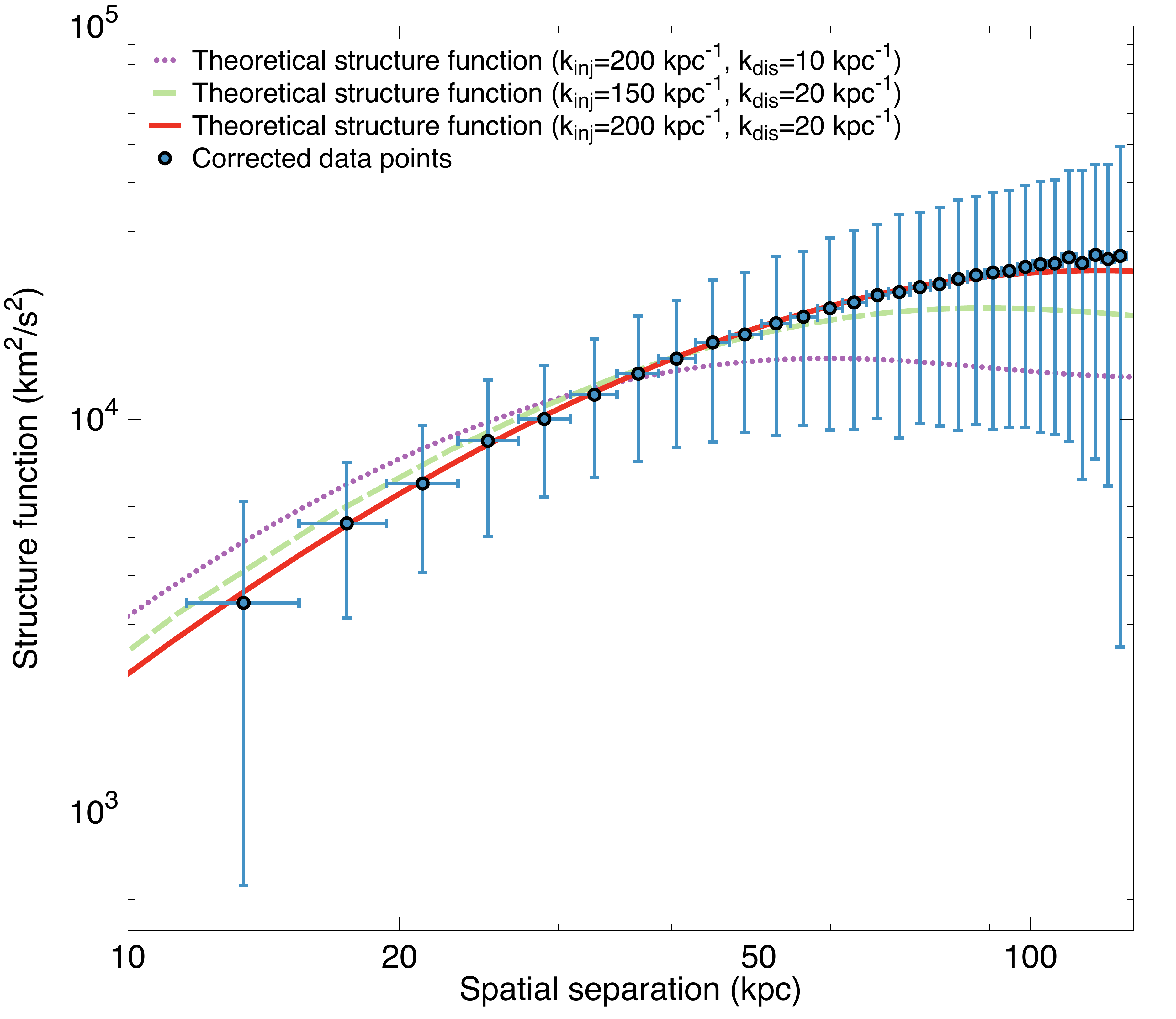}
\caption{Estimated structure function (km$^2$/s$^2$) averaged over 100 different observations of a velocity field generated with the same underlying turbulent velocity power spectrum as a function of the separation $s$ (kpc). (Left) Raw structure function recovered for $k_{\rm inj}=1/200$\,kpc$^{-1}$ and $k_{\rm dis}=1/20$\,kpc$^{-1}$. Different theoretical structure functions are also shown, the one associated to the run is given by red solid line. (Right) Same as left panel but the data points are corrected for the statistical bias and the binning projection effects (see CL19 for more information on the latter). Error bars indicate the  $\pm 1\sigma$ deviation within the 100 iterations.}
\label{fig:sfest}
\end{figure*}

\subsubsection{Post-processing of the data}

Due to limited statistics of the observation, pixels are binned into regions with a signal-to-noise (S/N) ratio $\textrm{S/N}=200$ ($\sim 40\,000$ counts per region) to reduce the statistical uncertainty on the measurements. To do so, we selected a spatial binning using an adapted Voronoï tessellation of the plane\footnote{\tiny{\url{https://www-astro.physics.ox.ac.uk/~mxc/software/\#binning}}} \citep{Cappellari2003Voronoi}, which provides $\sim 150$ regions over the X-IFU \textcolor{black}{FoV} ($\sim 20$ pixels per region, Figure~\ref{fig:turbmap} -- Bottom). This choice is motivated by the will to remain generic in our approach (this binning can be applied to any detector) and to provide round-shaped regions of constant S/N ratio, which ensure a faster convergence of the cosmic variance computation presented in CL19 than square regions.

The spectrum from each region is fitted using the input XSPEC model with an additional broadening component (\texttt{bapec}) to account for the effect of the turbulent velocities. A simultaneous fit of all the free parameters (temperature, abundance, redshift, velocity broadening, and norm) is performed. Results from the fits are excellent. No bias is visible on both parameters, and the statistical error distribution of the measured velocity shift, $\delta C_{\rm stat}$, is consistent with a centred Gaussian, of standard deviation $\sigma_{\textrm{stat}, C} = 34$\,km/s (Figure~\ref{fig:turbmap} -- Top right). This value of the statistical error was confirmed in every run, when using the same exposure and binning procedure (at constant $\beta$-model input). An example of binned input velocity map and the recovered output is provided Figure~\ref{fig:turbmap} (Bottom).

\section{Estimating the cosmic variance on X-IFU synthetic observations}
\label{sec:cosmic}

\subsection{General approach}

The previous E2E simulations are used as a test case to verify the formulas derived in Sect.~\ref{sec:estimation}. To do so, a computation of the structure function and its variance over a very large FoV would be required. This implies however to simulate large spatial grids with a refined mesh, which is rapidly computationally cumbersome (memory- and time-wise). We take advantage of the ergodicity assumption of the turbulence to simulate many independent velocity fields over the previous 2048 $\times$ 256 $\times$ 256 mesh, and average over these iterations to derive an estimation of the structure function. In practice, for a given choice of the turbulent velocity power spectrum (i.e. a subset of $\alpha$, $k_{\rm inj}$ and $k_{\rm dis}$), we create 100 different velocity fields, which are then observed using the previous E2E pipeline assuming a 100\,ks exposure time. We thus obtain -- for one choice of $P_{3D}$ -- 100 independent synthetic velocity maps of the same random process. 

In the case of the Coma cluster, we use as minimal dissipation scale of the turbulent velocity power spectrum $10$\,kpc, which represents a good compromise between current observational expectations \citep{Gaspari2013Turb} and future capabilities of the X-IFU (an X-IFU pixel size represents $\ell \sim 2$\,kpc at the Coma cluster redshift). Given the binned regions of our map ($\sim 5$ pixels of diameter) any scale larger than $10$\,kpc should be resolved. We know that large injection scales, a few hundred kpc up to 1\,Mpc, can occur due to merger event or subgroup accretion \citep[e.g.][]{Khatri2016Turb}. As we aim here at verifying our analytical formulation of the statistical error, we consider injection scales $\leq 200$\,kpc, which provide a likely description of the injection scale and keep the velocity field size within range of our available computation power. The slope of the turbulent power spectrum is fixed at $\alpha=-11/3$.

\subsection{Structure function estimation}

For a turbulent velocity power spectrum, $\overline{\textrm{SF}}$ can be estimated for each recovered velocity field through the previous formulas, using as separation $s$ the distances between the centres of binned regions $\mathcal{W}$.  For a Voronoï tessellation, the centres are found by taking the weighted barycentre with respect to the number of counts in each region $\mathcal{W}$. Given the homogeneous emission of the cluster toy model over the FoV, these points coincide in most of the cases with the geometrical centre of $\mathcal{W}$ (see the green crosses on Figure~\ref{fig:turbmap} -- Bottom right for an example). 

Though the emission profile of the cluster is the same, the non-constant region shape provided by the Voronoï tessellation creates slightly different spatial bins from one observation to the other. Hence, separations between regions do not follow a discrete mesh in each of the 100 observations of a given turbulent power spectrum. To compare the structure functions between the runs, we estimate SF on an \textit{a priori} grid of spatial separations, equal for each iteration, and with a step size of $\sim 5$\,kpc, which is approximately the equivalent radius of a Voronoï region. For instance, the regions with distances between 10 and 15\,kpc will be considered in the same bin to compute the value of $\overline{\textrm{SF}}$ in each run. The expected value of SF in the bin is then recovered by averaging over the 100 observations. The `true' value of spatial separation in the bin is taken by averaging all the real distances contained in the bin.
 
Examples of the estimated structure function for a given $P_{3D}$ are shown Figure~\ref{fig:sfest}, along with the corresponding $1\sigma$ deviation of each separation bin. The theoretical structure function associated with each 3D power spectrum is recovered through numerical integration of the analytical formula in a cylindrical frame ($r, \varphi, x$) along the line-of-sight $x$ by \citep{Zhuravleva2012SF}:
\begin{equation}
\textrm{SF}(s) = 4\pi \iint [1-J_0(2\pi k_r s)] P_{3D} (\underline{\vec{k}}) P_{\rho}^{~\vec{\theta}_{\rm eff}} (k_x) k_r dk_r dk_x
\end{equation}
where $J_0$ is the Bessel function of the first kind and $\vec{\theta}_{\rm eff}$ a fixed `effective' radius to compute the 1D power spectrum $P_{\rho}^{~\vec{\theta}}$. As the emissivity is not constant (see an example in Figure~\ref{fig:pow} -- Right), the structure function also varies depending on the regions considered to compute it (except in annular regions as the emission satisfies a spherical symmetry here). However, since $\epsilon$ is slowly-varying over the detector FoV in this case, only minor changes of the SF are expected ($\leq 5\%$ over the FoV). A good approximation of the observed structure function can be obtained by evaluating the previous formula on the annular radius corresponding to half of the equivalent radius of the detector. For the X-IFU this corresponds to $\theta_{\rm eff} \sim 34$\,kpc (1.24$^{\prime}$). 

As expected, the uncorrected values of the structure function are positively biased due to the statistical uncertainties in the measurements (Figure~\ref{fig:sfest} -- Left).  This bias can be corrected by subtracting the variance of the statistical error $\sigma_{\textrm{stat}, C}^2$ (see Equation~\ref{eq:esf}). Even with this correction, discrepancies of $\sim 5\%$ remain, which can be related to binning effects (see CL19). After subsequent correction (Figure~\ref{fig:sfest} -- Right), the average value of the structure function matches with the analytical structure function ($\leq 3\%$ in average on the relative difference between the simulations and the computed structure function over all separations). Remaining sources of error may be related to the number of iteration used here to recover the expected value of SF (100). Numerical effects related to the integration, the choice of $\theta_{\rm eff}$ and the binning may also create deviations.

\subsection{Structure function variance estimation}
\begin{figure*} [!t]
\centering
\includegraphics[width=0.49\textwidth]{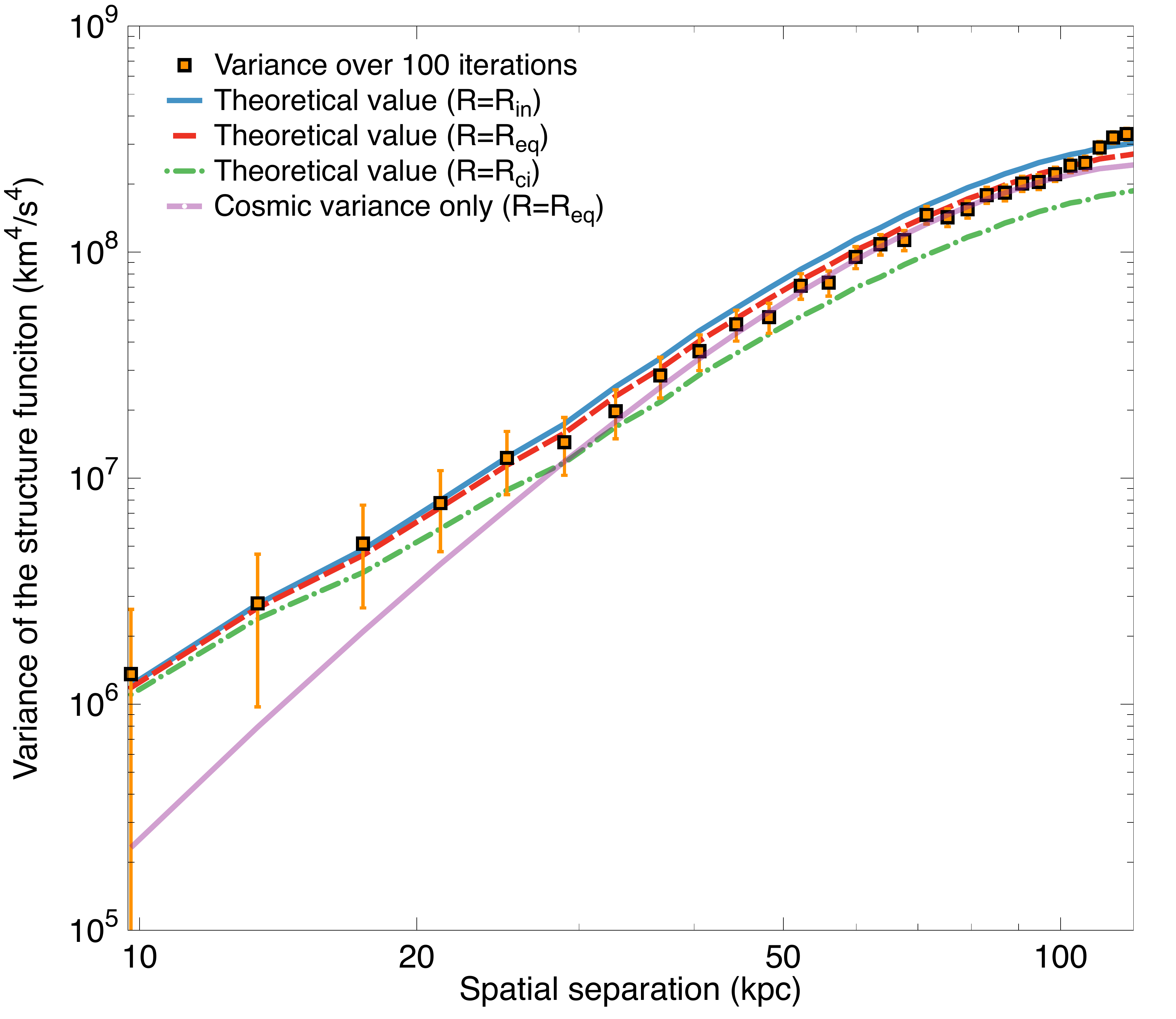}
\includegraphics[width=0.49\textwidth]{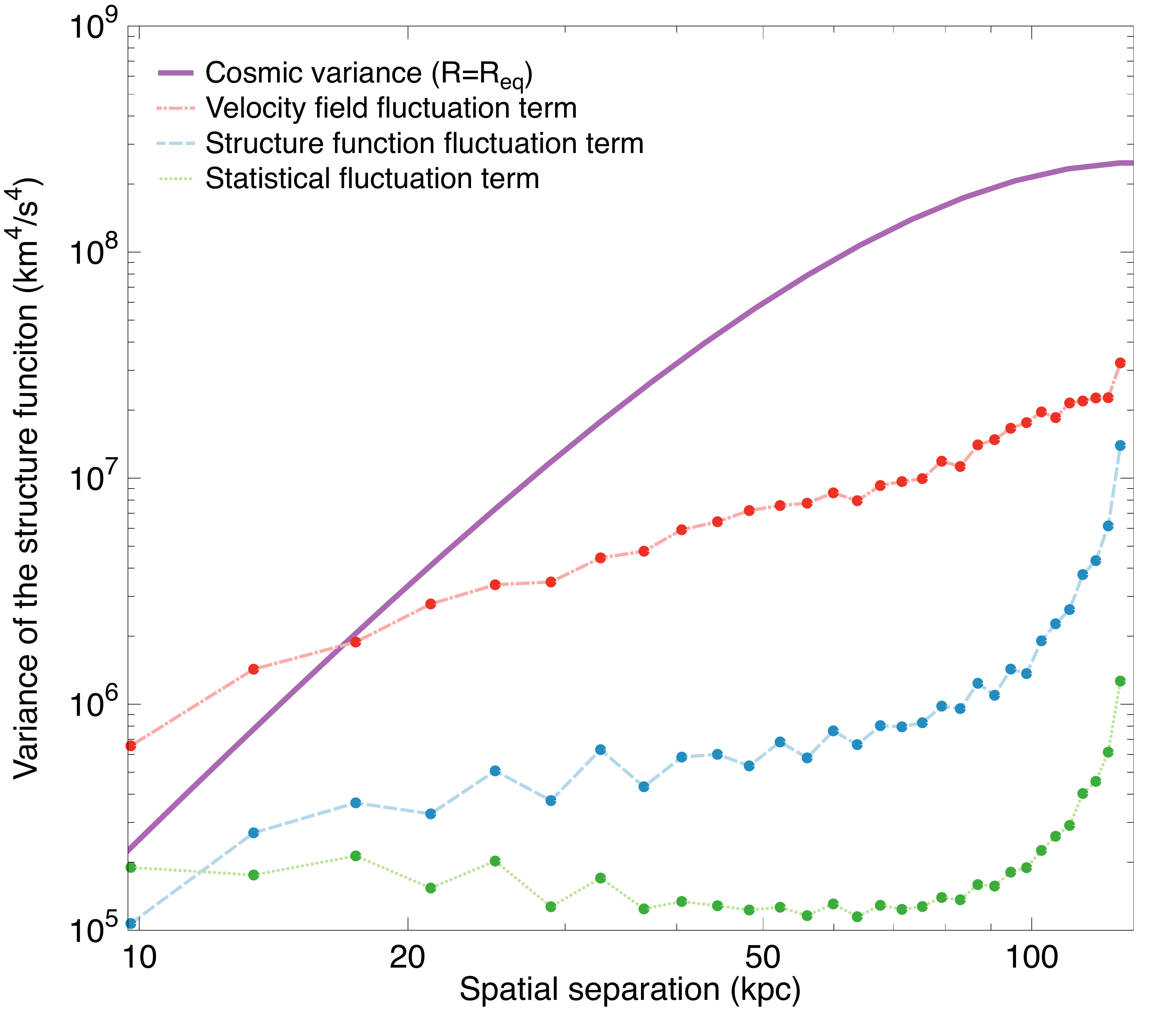}
\caption{(Left) Structure function variance (km$^4$/s$^4$) averaged over 100 observations of a velocity field generated with the same underlying turbulent power spectrum ($k_{\rm inj}=1/200$\,kpc$^{-1}$, $k_{\rm dis}=1/20$\,kpc$^{-1}$) and corresponding $\pm 1\sigma$ error bars  as a function of the separation $s$ (kpc). The comparison to the theoretical models derived from the simulation and the formulas by CL19 is shown for a circular FoV of $R_{\rm in}$ (blue solid), $R_{\rm eq}$ (red dashed) or $R_{\rm ci}$ (green dash-dotted). The sheer contribution of the cosmic variance without statistical terms for $R=R_{\rm eq}$ is given in light purple. (Right). Error contributions in the total variance for the previous case for $R=R_{\rm eq}$ when $\sigma_{\textrm{stat}, C}=34$\,km/s. The data points for the statistical and structure function fluctuation term are derived from analytical formulas, while the velocity field fluctuation term is computed using the 100 iterations (see text).}
\label{fig:sferr}
\end{figure*} 

Similarly, the variance of the structure function over the runs can be compared to Equation~\ref{eq:varsf}. The geometrical terms, $N_{nei}(s)$ and $N_p(s)$, are derived from the binning map. $\textrm{D}$ can be computed through $\overline{\textrm{D}}$, unbiased for a centred Gaussian statistical noise. Its variance however, is biased such that (see Appendix~\ref{app:vard}):
\begin{equation}
\textrm{Var}[\overline{\textrm{D}}(s)] = \textrm{Var}[\textrm{D}(s)] + \frac{2 \sigma_{\textrm{stat}, C}^2 N_{nei}(s)}{N_p(s)}
\label{eq:vard}
\end{equation} 
Var[$\textrm{D}(s)$] is therefore estimated through Var[$\overline{\textrm{D}}(s)$] over the 100 simulations, and corrected of its bias. The intrinsic cosmic variance is obtained using the formulas provided in CL19. To do so, a specific circular FoV of the instrument and a pixel size (or bin size) are needed. We consider here that bins are well described by disks of the same diameter as the Voronoï regions. \textcolor{black}{Similarly} the hexagonal detector is approximated by an equivalent disk. Three different options were considered for its radius $R$:
\begin{itemize}
\item `Equivalent' radius $R_{\rm eq}$. If $\mathcal{S}_\mathcal{A}$ is the total area of the detector, $R_{\rm eq} = \sqrt{\mathcal{S}_\mathcal{A}/\pi}$ = 67.6\,kpc = $149.4^{\prime \prime}$.
\item Radius of the inscribed circle, $R_{\rm in}$ = 63.9\,kpc = $141.2^{\prime \prime}$.
\item Radius of the circumscribed circle, $R_{\rm ci}$ = 82.1\,kpc = $181.4^{\prime \prime}$.
\end{itemize}

Figure~\ref{fig:sferr} (Left) shows the comparison for different values of $R$, and the case without statistical corrections. For large separations, the statistical terms presented in Equation~(\ref{eq:varsf}) have little effects, but must be accounted for when considering smaller separations. Also, despite the slight differences between the considered radii, changes in the analytical values of the variance of a factor 2 are observed when statistics are included. Simulation points are comprised between the $R=R_{\rm ci}$ and $R=R_{\rm in}$ curves, and all three curves show a good agreement within error bars. Deviations between the simulated data and predicted errors is lower than 20\% for all separations (i.e. less than 10\% in standard deviation) with $R_{\rm eq}$ providing the best results ($10\%$ in variance, hence $\leq 5\%$ in standard deviation). 

These curves accurately recover the shape of the expected variance, but show a consistent deviation at large separations, which can be related to two distinct effects. On the one hand, these separations are sampled only a handful of times within a single X-IFU pointing (i.e. one or two regions per iteration are separated by a detector diameter), thus creating a simulation-related sample variance in the data. On the other hand, the circular FoV approximation reaches a limit for separations of the same order (or higher) than $R_{\rm eq}$. Smaller deviations could also be caused by box-size effects of the turbulent velocity grid.

Other factors can explain the remaining deviations between the theoretical curves and the simulations:
\begin{itemize}
\item The computation of the analytical error formulas involve complex numerical integrals, which can account for errors in the variance around 1 to 2\%.
\item The E2E simulation sample used in the simulations (100) is sufficiently large to have a good estimate of the average, but may be insufficient for an accurate value of the variance, up to a level of 10\%. Results with 250 iterations indicate a better agreement with the theoretical error formulas (improvement of a few \%), suggesting that part of the deviations are consistent with a small size of the sample. To reduce this contribution within numerical errors (i.e. below the 2\% threshold) at least 2500 iterations are required. Such an increase would however render the computational time unreasonable ($\sim 1$ week of computational time for 250 iterations for a given $P_{3D}$ with our current set-up).
\item The approximation of the X-IFU FoV with a disk. A better description of the detector geometry in the computation of the intrinsic variance component could reduce this component, especially for separations larger than the detector equivalent diameter.
\item The bin shape used for the numerical computation, which assumes a uniform tessellation of the detector. Voronoï binning was chosen to represent a more generic case than simple square regions. It does not, however, verify a uniform tessellation. Test runs performed using square groups of pixels show similar results in the comparison, suggesting little impact of this particular effect on our previous results.
\end{itemize}

This analysis was extended to other turbulent velocity power spectra, with $k_{\rm inj}$ within 100 and 200\,kpc, and $k_{\rm dis}$ from 10\,kpc to 30\,kpc, showing similar results. 
\begin{figure*} [!t]
\centering
\includegraphics[width=0.49\textwidth]{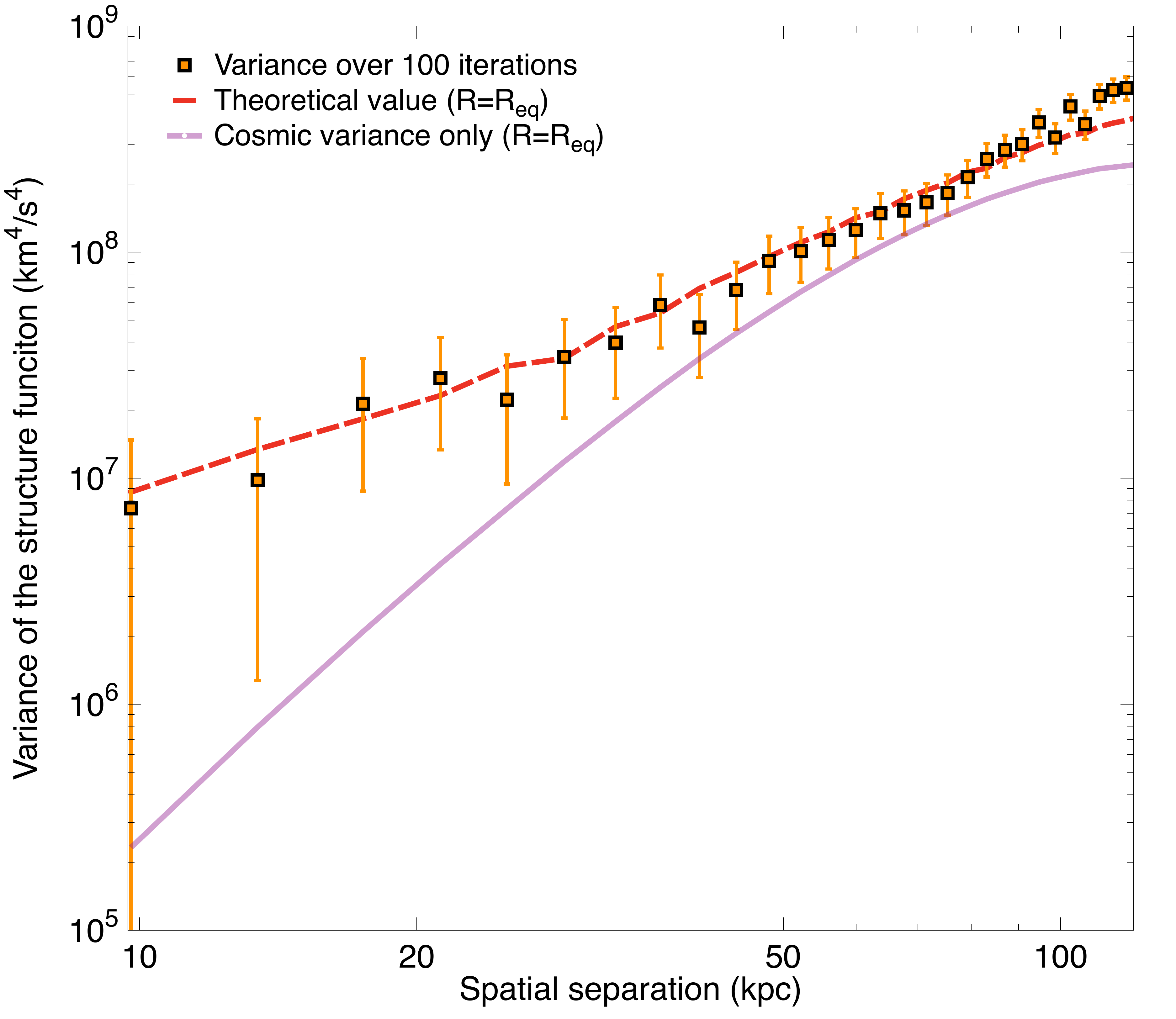}
\includegraphics[width=0.49\textwidth]{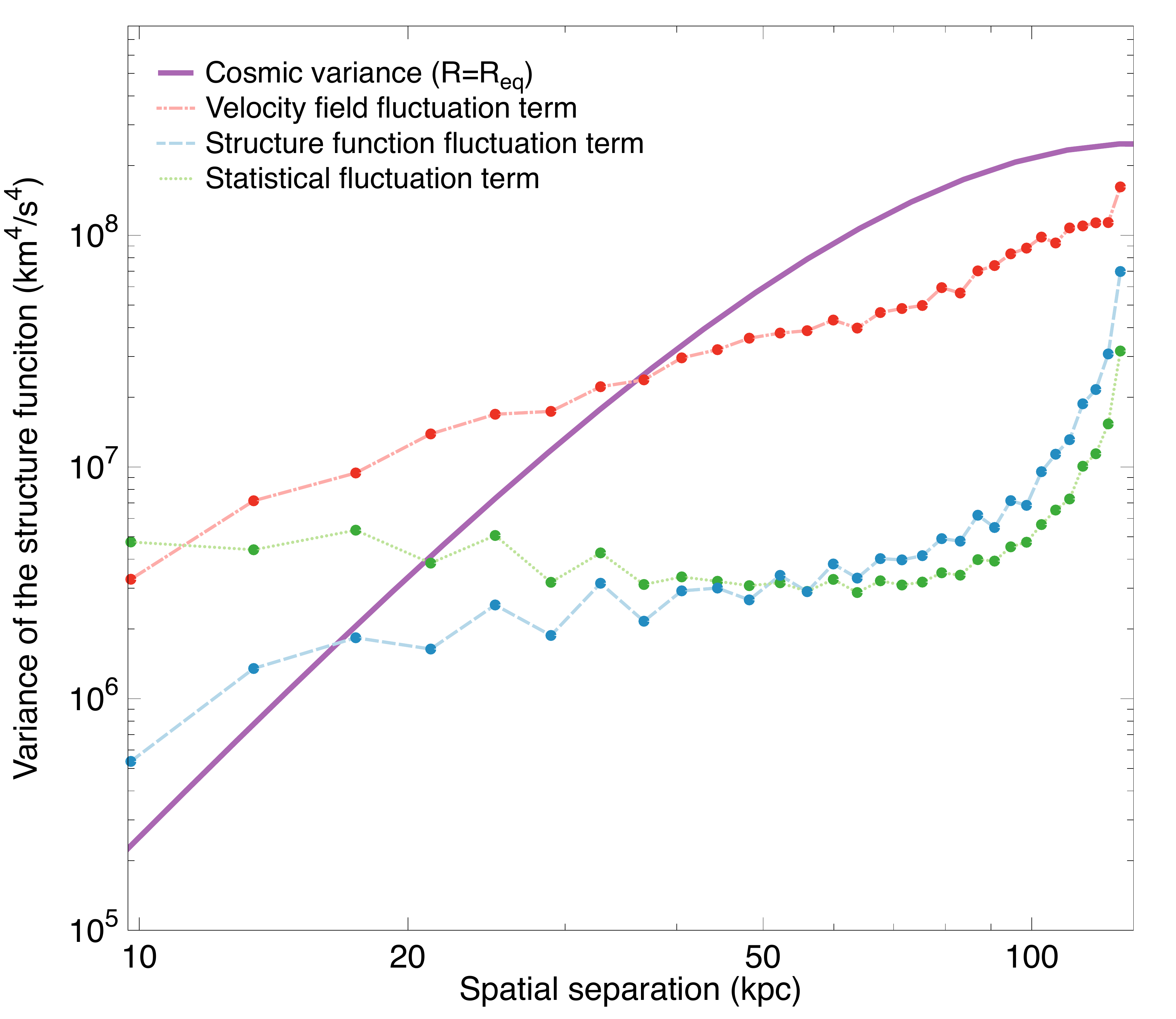}
\caption{Same as Figure~\ref{fig:sferr} for 20\,ks observations using the same binning map. Only the case $R=R_{\rm eq}$ is shown.}
\label{fig:sferr2}
\end{figure*} 

\section{Implications of error formulas}
\label{sec:discuss}

\subsection{Validation of error formulas and relative contribution}

Figure~\ref{fig:sferr} (Right) shows the contribution of the sources of error for a given $P_{3D}$. Among the statistical error contributions, computed in each bin, the velocity field fluctuation term dominates the other two and shows a monotonous trend. The structure function fluctuation term is lower than the previous contributor, but becomes comparable at high $s$, where the number of regions used to compute SF is small. Finally, as expected, the purely statistical term is the smallest and can in most cases be neglected. Its contribution is mostly on small separations and minimal on average separations, where $N_p$ is highest. We note however that its value becomes comparable to the cosmic variance for small $s$. 

For  $\sigma_{\textrm{stat}, C}=34\,$km/s, used here,  the intrinsic cosmic variance term dominates over most of the separations ($s \geq 30/40$\,kpc). This result holds for all the $P_{3D}$ tested. As suggested by Figure~\ref{fig:sferr} (Left), the correction included by the statistical terms in the estimation of the total variance is mainly visible at small separations, validating the formulas out to $\sim 50$\,kpc. Testing the previous formulas for high $s$ requires to decrease the contribution of the cosmic variance or to increase the other statistical contributors to enhance error terms for high $s$. All things being equal, these terms scale with $\sigma_{\textrm{stat}, C}^2$ or $\sigma_{\textrm{stat}, C}^4$, and become dominant when $\sigma_{\textrm{stat}, C} \geq 100$\,km/s. This can thus be achieved using larger pointings with a similar binning and exposure (to reduce the cosmic variance), or by artificially increasing the statistical error by for instance considering shallower exposures. The latter was considered, by reducing the previous runs to 20\,ks exposures, and provides a good agreement with the theoretical formulas (see Figure~\ref{fig:sferr2}). 

\subsection{Practical estimation of errors}

 As shown in CL19, an estimation of the cosmic variance can be obtained numerically with several approximations on the detector geometry. The statistical terms however are intrinsically related to the observational set-up. We provide here some solutions to estimate the quantities involved in Equations~\ref{eq:esf} and \ref{eq:varsf} for a specific instrumental configuration.

The values of $N_p(s)$ and $N_{nei}(s)$ are the simplest to derive, as they are related to the binning and geometry of the detector, and can be determined analytically with high accuracy.  $\sigma_{\textrm{stat}, C}$ will be a direct output of the observation. However, an \textit{a priori} estimation of the statistical error of an observation, which could be used to forecast structure function errors before an observation (e.g. to optimise the exposure strategy) is more challenging. Crude estimations can be derived using a simple Poissonian approach using the flux of the source and the exposure time. Provided E2E simulations of the instrument become sufficiently representative, a promising solution could be to derive $\sigma_{\textrm{stat}, C}$ numerically (a single pointing is required). 

The variance of $\textrm{D}$ is by far the most challenging term to estimate, as it requires multiple observations. In our simulations, the ergodicity assumption simplifies these computations, as accurate estimations of the previous terms can be obtained by averaging over a large number of iterations. This approach does not hold in flight, as multiple realisations of the same turbulent field are unlikely to be met (statistically speaking), even assuming a self-similar behaviour of the turbulence between clusters. Even if single pointings of a larger mapping of the same object are performed, an accurate computation of the variance cannot be achieved unless the emission profile does not vary across the FoV, which only concerns a handful of clusters (at a zeroth order approximation).  A solution is once again to use numerical simulations. An idea is to find analytical formulas, similarly to CL19, and use them to obtain a fast numerical estimation of Var[$\textrm{D}(s)$] for a given power spectrum assumption. \textcolor{black}{As this remains to be investigated, dedicated MC simulations remain at the moment the other possible solution. Although relying once again on an iterative method, these simulations do not require a full E2E approach, and can be performed without any prior on the instrument except its geometry (implying large gains in computational time).}

\subsection{Towards optimising observation strategies}

Under the current assumption of 100\,ks pointings, we expect statistical errors on X-IFU observations to be low and well constrained. Hence, over a single pointing,  an accurate knowledge of the intrinsic cosmic variance will be sufficient to provide an error estimation of the SF at large separations (Figure~\ref{fig:sferr}), and thereby to investigate turbulence at large scales (i.e. injection scale). Statistical terms can in this case be neglected in the error computation. On the contrary, when measuring the structure function for small separations (i.e. to probe dissipation scales), a good estimation of all the statistical terms is paramount to ensure a proper error description. Depending on the science objectives of the X-IFU (large- or small-scale turbulence investigations), the observational strategy can then be optimised. 

For a constant exposure time, several options can be explored. Deep pointings would provide a constant value of the cosmic variance, but significantly reduce $\sigma_{\textrm{stat}, C}$, which may be interesting to explore small separations. On the contrary, since statistical errors are negligible at high $s$, multiple shallower pointings could be used to explore larger separations (notably larger than the detector) while reducing the cosmic variance. An optimal point where all errors are comparable across the separations could also be used. The formulas derived here and in CL19 demonstrate that accurate error estimations of the SF can be found by numerical integration. This approach is complementary to studies of turbulence limited through an iterative approach (e.g. 100 simulations of a large grids and associated errors), which can now be reduced to a handful of simulations. 

\section{Conclusion}

With improvements in high-resolution spatially resolved X-ray spectroscopy, \textcolor{black}{measurements of line shifts, line broadening and structure functions} will provide new insight on the turbulence at play within the ICM. In this paper, we addressed the challenge of computing these diagnostics and estimating their errors, related to both cosmic variance and measurement uncertainties. Specifically, this work extends the approach started in our companion paper I, which derives a formulation for the cosmic variance, and adds the contribution of finite statistics in the observations. \emph{All the formulas presented here should thus be coupled to those from Clerc et al. (2019, sub.) on the intrinsic cosmic variance.}

We found that all the estimators, notably those of the structure function and its variance over various observations, are biased by the statistics of the measurements. For the variance of the structure function, these biases can be divided \textcolor{black}{into multiple contributors, which add to the intrinsic cosmic variance: a purely statistical contributor, which scales with the squared variance of the statistical uncertainty ($\sigma_{\textrm{stat}, C}^4$), and a contribution related to the intrinsic nature of the velocity field, which scales with $\sigma_{\textrm{stat}, C}^2$ and depends on the spatial binning, the detector geometry, and the scales of the turbulence (injection and dissipation). We divided the latter into two terms in the case of Gaussian error field,  a structure function fluctuation term related to the turbulent velocity power spectrum, and a velocity field fluctuation term, related to the scales of the turbulence within the FoV.} \emph{The equations derived in Sect.~\ref{sec:estimation} are generic and valid for any instrument with measurement uncertainties}. 

A specific application to the X-IFU instrument was performed to demonstrate the validity of these formulas in the framework of the future mission \textsl{Athena}. For this test, turbulent velocity fields of a toy model Coma-like cluster were generated for different underlying turbulent power spectra (here for different injection and dissipation scales of a Kolmogorov power spectrum), and used as inputs for synthetic observations with the X-IFU E2E simulator SIXTE. When such observations are averaged over a large number of different realisations of the velocity field (assuming ergodicity) and corrected of their biases, our results show excellent agreement with the analytical values, with relative errors below $3\%$ over the spatial scales investigated. A comparison of the variance of the estimated structure function using the analytical error formulas presented here and in Clerc et al. (2019, sub.) also provides accurate result (better than 10\% in variance, hence $\leq 5\%$ in standard deviation).

Results presented here demonstrate that we can provide accurate estimations of the total variance of the structure function. For typical X-IFU observations of 100\,ks, statistical terms in the structure function errors can be neglected for large spatial separations ($s \geq 70$\,kpc), but are required to investigate smaller separations ($s \leq 30$\,kpc). Depending on the science case, efforts can therefore be directed into reducing one or several of the error terms specifically, or into optimising the observational strategy (exposure and spatial map) and the spatial binning (region size and shape). A dedicated analysis of these optimisations, taking advantage of the fast computations enabled by the formalism proposed here, will be discussed in a forthcoming study.  Measurements remain all the same a challenging objective, especially when systematics or the physics of the ICM (e.g. AGN, shocks) are considered. New results will thus also rely on alternative diagnostics, such as line non-Gaussianity, achievable through the spectral resolution of the X-IFU.

\small{\paragraph{\textit{Acknowledgements}}
The authors would like to thank the anonymous referee for the suggestions and helpful comments. We thank D.~Barret as well for the fruitful discussions on the paper. Finally, the authors also acknowledge the SIXTE development team at ECAP, CNES, IRAP, and IFCA (in particular T.~Dauser and J.~Wilms) for the development and distribution of state-of-the-art simulation tools for the X-IFU. 

%


\bibliographystyle{aa}
\bibliography{paper}

\appendix

\section{Expected average and variance of the line diagnostics with finite statistics}
\label{app:computation}

We derive here the expected values for the average and variance of the structure function estimator. For these computations, we assume ergodicity and isotropy of the turbulence processes. Under these assumptions, the expectation $\mathbb{E}$  over the entire space is equal to the average on a single point when considering an infinite number of realisation of the turbulent process. The expectation operator $\mathbb{E}$ is linear, such that for a given random variable $X$, $\textrm{Var}[X]=\mathbb{E}[X^2] - \mathbb{E}[X]^2$. The Var operator indicates the variance over different observations or subset of observations, for instance in the case of multiple pointings or regions. It does not include any systematic or projection-related effect. 

We remind that when the random variable $\delta C_{\rm stat}$ describing the statistical error is a centred Gaussian of standard deviation $\sigma_{\textrm{stat}, C}$, and independent from the underlying turbulent velocity field. Its first four moments for any point ${\vec{x}}$ in a 2D space over an infinite number of random realisations are given by
\begin{align*}
\mathbb{E}[\delta C_{\textrm{stat}}(\vec{x})] &= 0 \\
\mathbb{E}[\delta C_{\textrm{stat}}(\vec{x})^2] &= \sigma_{\textrm{stat}, C}^2  \\
\mathbb{E}[\delta C_{\textrm{stat}}(\vec{x})^3] &= 0 \\
\mathbb{E}[\delta C_{\textrm{stat}}(\vec{x})^4] &= 3\sigma_{\textrm{stat}, C}^4 
\end{align*}

\subsection{Line shift and line broadening}
\label{subapp:line}

\paragraph{Line shift}\mbox{}\\

\noindent In the case of line shift, if we assume that the turbulent velocity field is centred, the expected value of $\overline{C}$ averaged over multiple different realisation of the velocity field over a point or region $\vec{x}$ is simply
\begin{align*}
\mathbb{E}[\overline{C}(\vec{x})] &= \mathbb{E}[C(\vec{x})] + \mathbb{E}[\delta C(\vec{x})_{\rm stat}]  \\
&=0
\end{align*}
Similarly, the variance is given by
\begin{align*}
\textrm{Var}[\overline{C(\vec{x})}] & = \mathbb{E}[\overline{C}^2(\vec{x})] - \mathbb{E}[\overline{C}(\vec{x})]^2 \\
&= \mathbb{E}[C^2(\vec{x})] -  \mathbb{E}[C(\vec{x})]^2  + 2\mathbb{E}[\delta C(\vec{x})_{\rm stat}] \mathbb{E}[C(\vec{x})] \\
&~~~ + \mathbb{E}[\delta C(\vec{x})_{\rm stat}^2] \\
&= \textrm{Var}[C(\vec{x})] + \sigma_{\textrm{stat}, C}^2
\end{align*}

\paragraph{Line broadening}\mbox{}\\

\noindent Using a similar approach as the previous case, we find that 
\begin{align*}
\mathbb{E}[\overline{\tilde{S}^2}(\vec{x})] &= \mathbb{E}[(\tilde{S}(\vec{x})+\delta \tilde{S}(\vec{x})_{\rm stat})^2] \\
&=\mathbb{E}[\tilde{S}^2(\vec{x})] + 2\mathbb{E}[\tilde{S}(\vec{x})]\mathbb{E}[\delta \tilde{S}(\vec{x})_{\rm stat}]+\mathbb{E}[\delta \tilde{S}(\vec{x})_{\rm stat}^2] \\
&=\mathbb{E}[\tilde{S}^2(\vec{x})] + \sigma_{\textrm{stat}, \tilde{S}}^2
\end{align*}
where $\mathbb{E}[\tilde{S}^2(\vec{x})] = \sigma_{\rm turb}^2 - \textrm{Var}[C(\vec{x})]$ (see CL19). For the variance:
\begin{align*}
\textrm{Var}[\overline{\tilde{S}^2}(\vec{x})] &= \mathbb{E}[(\tilde{S}(\vec{x})+\delta \tilde{S}(\vec{x})_{\rm stat})^4] - \mathbb{E}[(\tilde{S}(\vec{x})+\delta \tilde{S}(\vec{x})_{\rm stat})^2]^2 \\
&=\mathbb{E}[\tilde{S}^4(\vec{x})] + 6\mathbb{E}[\tilde{S}^2(\vec{x})]\mathbb{E}[\delta \tilde{S}(\vec{x})_{\rm stat}^2]+\mathbb{E}[\delta \tilde{S}(\vec{x})_{\rm stat}^4] \\
&~~~ + 4 \mathbb{E}[\tilde{S}^3(\vec{x})]\mathbb{E}[\delta \tilde{S}(\vec{x})_{\rm stat}] + 4 \mathbb{E}[\delta \tilde{S}(\vec{x})_{\rm stat}^3]\mathbb{E}[\tilde{S}(\vec{x})] \\ 
&~~~ - \mathbb{E}[\tilde{S}^2(\vec{x})]^2 - 2 \mathbb{E}[\tilde{S}^2(\vec{x})] \mathbb{E}[\delta \tilde{S}(\vec{x})_{\rm stat}^2] - \mathbb{E}[\delta \tilde{S}(\vec{x})_{\rm stat}^2]^2 \\
&= \textrm{Var}[\tilde{S}(\vec{x})^2] + 4 \mathbb{E}[\tilde{S}(\vec{x})^2] \sigma_{\textrm{stat}, \tilde{S}}^2 + 2 \sigma_{\textrm{stat}, \tilde{S}}^4
\end{align*}

\subsection{Neighbours and detector tessellation}
\label{subapp:nei}

\noindent For future computations, the geometry of the detector will need to be accounted for. We define here several quantities used throughout the rest of the study. \vspace{-0.2cm}

\paragraph{Definition A:} Let it be a given tessellation $\mathcal{T}$ of the observed space (here $\mathbb{R}^2$) corresponding to the pixel configuration of the instrument or the binned regions of an observation. For a given point ${\vec{x}}$, centre of one of these regions, we define $N_{nei}(s)$ as the number of tessellated regions in its vicinity whose centre are at a distance $s$ from ${\vec{x}}$ (in the sense of the Euclidian norm). \vspace{-0.2cm}

\paragraph{Definition B:} Let us assume an infinite tessellation $\mathcal{T}$ of the observed space (neglecting border effects) and a given $\vec{p} = ({\vec{x}}, \vec{y}) \in \mathbb{R}^4$ with $||{\vec{x}}, \vec{y}||_2=s$.  $\mathcal{V}_{{\vec{x}}, \vec{y}} \subset \mathbb{R}^4$ is the subset that contains all the pairs $\vec{q}=({\vec{w}}, {\vec{z}}) \in \mathbb{R}^4$ such that $\vec{q} \neq \textbf{p}$, $||{\vec{w}}, {\vec{z}}||_2 =s$, with ${\vec{w}} = {\vec{x}}, {\vec{z}} \neq \vec{y}$ or ${\vec{w}} \neq {\vec{x}}, {\vec{z}} = \vec{y}$ strictly. \vspace{-0.2cm}

\paragraph{Lemma A:} For a given $\vec{p} = ({\vec{x}}, \vec{y}) \in \mathbb{R}^2$, $\mathcal{V}_{{\vec{x}}, \vec{y}}$ is of cardinal $N_{p}(s) (N_{nei}(s) -1)$.\\

\noindent \textbf{Proof:} The total number of pairs $\vec{p}=({\vec{x}}, \vec{y}$) with $||{\vec{x}}, \vec{y}||_2=s$ is $2 N_{p}(s)$. Let us now work on the `halved' space where no permutations between ${\vec{x}}$ and $\vec{y}$ are considered. For two pairs $\vec{p}, \vec{q}$ of points, the total number of different combinations is given by:
$$C_2^{N_{p}(s)} = \frac{N_{p}(s) (N_{p}(s)-1)}{2}$$ 

If a pair $\vec{p}$ is chosen, we have $N_{p}(s)$ different combinations to choose from. However once this pair is selected the choice of $\vec{q} \not \in \mathcal{V}_{{\vec{x}}, \vec{y}}$ is given by $(N_{p}(s) - 2(N_{nei}(s)-1) -1)$. Since the permutation between the selection counts elements twice, the number of elements inside  $\mathcal{V}_{{\vec{x}}, \vec{y}}$, that is $\#\mathcal{V}_{{\vec{x}}, \vec{y}}$ is 
\begin{align*}
\#\mathcal{V}_{{\vec{x}}, \vec{y}} &= \frac{N_{p}(s) (N_{p}(s)-1)}{2} - \frac{N_{p}(s)(N_{p}(s) - 2(N_{nei}(s)-1) -1)}{2} \\
&= N_{p}(s) (N_{nei}(s)-1)
\end{align*}

\subsection{Expected average of the structure function}
\label{subapp:esf}

\noindent The definition of the estimator accounting for the statistical uncertainties can be written as
\begin{align*}
\mathbb{E}[\overline{\textrm{SF}}(s)] &= \sum_{({\vec{x}}, \vec{y}) \in \mathbf{\tilde{S}}_s}  \frac{\mathbb{E}[ (C({\vec{x}}) + \delta C({\vec{x}})_{\rm stat} - C(\vec{y})- \delta C(\vec{y})_{\rm stat})^2 ]}{N_{p(s)}} \\
&= \frac{1}{N_p(s)} \sum_{({\vec{x}}, \vec{y}) \in \mathbf{\tilde{S}}_s}  \mathbb{E}[ (C({\vec{x}}) - C(\vec{y}))^2 + \delta C({\vec{x}})_{\rm stat}^2  \\
& ~~~ + \delta C(\vec{y})_{\rm stat}^2 - 2 \delta C({\vec{x}})_{\rm stat}\delta C(\vec{y})_{\rm stat} \\
& ~~~ - 2 (C({\vec{x}}) - C(\vec{y})) (\delta C({\vec{x}})_{\rm stat} - \delta C(\vec{y})_{\rm stat})] 
\end{align*}
Each of these six terms can be evaluated independently. \\

\noindent \textcircled{\tiny{1}}
By definition of the structure function we have that:
\begin{align*}
\frac{1}{N_p(s)} \sum_{({\vec{x}}, \vec{y}) \in \mathbf{\tilde{S}}_s} \mathbb{E}[(C({\vec{x}}) - C(\vec{y}))^2] = \textrm{SF}(s)
\end{align*}

\noindent \textcircled{\tiny{2}} / \textcircled{\tiny{3}} Using the moments of the central Gaussian field:
\begin{align*}
\frac{1}{N_p(s)} \sum_{({\vec{x}}, \vec{y}) \in \mathbf{\tilde{S}}_s} \mathbb{E}[\delta C(\vec{y})_{\rm stat}^2] &= \sigma_{\textrm{stat}, C}^2
\end{align*}

\noindent \textcircled{\tiny{4}} + \textcircled{\tiny{5}}
\begin{align*}
\begin{split}
\frac{2}{N_p(s)} & \sum_{({\vec{x}}, \vec{y}) \in \mathbf{\tilde{S}}_s} \mathbb{E}[(C({\vec{x}}) - C(\vec{y})) (\delta C({\vec{x}})_{\rm stat} - \delta C(\vec{y})_{\rm stat})] \\
&= \frac{2}{N_p(s)} \sum_{({\vec{x}}, \vec{y}) \in \mathbf{\tilde{S}}_s} (\mathbb{E}[C({\vec{x}})\delta C({\vec{x}})_{\rm stat}] - \mathbb{E}[C({\vec{x}})\delta C(\vec{y})_{\rm stat}] \\
 &~~~ - \mathbb{E}[C(\vec{y})\delta C({\vec{x}})_{\rm stat}]+\mathbb{E}[C(\vec{y})\delta C(\vec{y})_{\rm stat}])) \\
\end{split}
\end{align*}
Since the fields are independent from the errors, each term of the previous equation can be decoupled (e.g. $\mathbb{E}[C({\vec{x}})\delta C(\vec{y})_{\rm stat}] = \mathbb{E}[C({\vec{x}})] \mathbb{E}[\delta C(\vec{y})_{\rm stat}]$). The error distribution being centred, the overall sum is 0 for each point of $\mathbf{\tilde{S}}_s$. \\

\noindent \textcircled{\tiny{6}}
By definition of $\mathbf{\tilde{S}}_s$, ${\vec{x}} \neq \vec{y}$, similarly to the previous case
\begin{align*}
\frac{2}{N_p(s)} \sum_{({\vec{x}}, \vec{y}) \in \mathbf{\tilde{S}}_s} \mathbb{E}[\delta C({\vec{x}})_{\rm stat}\delta C(\vec{y})_{\rm stat}]&=0
\end{align*}
The final result is obtained by summing the six terms:
\begin{align*}
\mathbb{E}[\overline{\textrm{SF}}(s)] = \textrm{SF}(s)+2 \sigma_{\textrm{stat}, C}^2
\end{align*}

\subsection{Expected variance of the structure function}

\noindent We take a specific interest here to the variance of the previous estimators. To compute the expected variance of $\overline{\textrm{SF}}$, we proceed in the same way as in Sect.~\ref{subapp:esf}. The variance of the estimator is defined as 
\begin{align*}
\mathrm{Var}&[\overline{\textrm{SF}}(s)] \\
&~~~ = \mathrm{Var}[\sum_{({\vec{x}}, \vec{y}) \in \mathbf{\tilde{S}}_s} \frac{[(C({\vec{x}})-C(\vec{y}))+\delta C({\vec{x}})_{\rm stat} - \delta C(\vec{y})_{\rm stat}]^2}{N_p(s)}] \\
&~~~ =  \frac{1}{N_p(s)^2} \mathrm{Var}[\sum_{({\vec{x}}, \vec{y}) \in \mathbf{\tilde{S}}_s}  \zeta_{\vec{x},\vec{y}}^2]
\end{align*}

\noindent where we define $\zeta_{{\vec{x}}, \vec{y}} = (C({\vec{x}})-C(\vec{y}))+\delta C({\vec{x}})_{\rm stat} - \delta C(\vec{y})_{\rm stat}$ and $\zeta_{0,{\vec{x}}, \vec{y}}= C({\vec{x}})-C(\vec{y})$. Since the variables are not necessarily independent ($({\vec{x}}, \vec{y})$ may be two-by-two different in $\mathbb{R}^4$ but they may share the same ${\vec{x}}$ or $\vec{y}$), the previous expression becomes

\begin{align*}
\mathrm{Var}[\overline{\textrm{SF}}(s)] &=  \frac{1}{N_p(s)^2} ( \sum_{({\vec{x}}, \vec{y}) \in \mathbf{\tilde{S}}_s} \mathrm{Var}[\zeta_{{\vec{x}}, \vec{y}}^2] \\ 
&~~~ +\sum_{\substack{({\vec{x}}, \vec{y}), ({\vec{x}^{\prime}}, {\vec{y}^{\prime}}) \in \mathbf{\tilde{S}}_s \\ ({\vec{x}}, \vec{y}) \neq ({\vec{x}^{\prime}}, {\vec{y}^{\prime}})}} \textrm{Cov} [\zeta_{{\vec{x}}, \vec{y}}^2, \zeta_{{\vec{x}^{\prime}}, {\vec{y}^{\prime}}}^2] )\\
\end{align*}

\noindent The first term represents the expected variance of the parameters (non-zero for any given couple of $({\vec{x}}, \vec{y})$) and the second term the covariance of the parameters. The covariance term is non-zero exclusively when the two pairs of points share one given point (i.e. ${\vec{x}}=\vec{x}^{\prime}$ or $\vec{y}={\vec{y}^{\prime}}$).

\paragraph{Variance term}\mbox{}\\
\begin{align*}
\frac{1}{N_p(s)^2} \sum_{({\vec{x}}, \vec{y}) \in \mathbf{\tilde{S}}_s} \mathrm{Var}[\zeta_{{\vec{x}}, \vec{y}}^2] &= \frac{1}{N_p(s)^2} \sum_{({\vec{x}}, \vec{y}) \in \mathbf{\tilde{S}}_s} \mathrm{Var}[ (C({\vec{x}})-C(\vec{y}))^2  \\
&~~~ + \delta C({\vec{x}})_{\rm stat}^2 + \delta C(\vec{y})_{\rm stat}^2 \\
&~~~ - 2 \delta C({\vec{x}})_{\rm stat} \delta C(\vec{y})_{\rm stat} \\
&~~~ + 2 (C({\vec{x}})-C(\vec{y})) (\delta C({\vec{x}})_{\rm stat} - \delta C(\vec{y})_{\rm stat})]
\end{align*}

\noindent \textcircled{\tiny{1}}
\begin{align*}
\frac{1}{N_p(s)^2} \sum_{({\vec{x}}, \vec{y}) \in \mathbf{\tilde{S}}_s} \mathrm{Var}[(C({\vec{x}})-C(\vec{y}))^2] = \frac{1}{N_p(s)^2} \sum_{({\vec{x}}, \vec{y}) \in \mathbf{\tilde{S}}_s} \mathrm{Var}[\zeta_{0, {\vec{x}}, \vec{y}}^2]
\end{align*}

\noindent \textcircled{\tiny{2}} / \textcircled{\tiny{3}}
\begin{align*}
\mathrm{Var}[\delta C({\vec{x}})_{\rm stat}^2] &= ( \mathbb{E}[\delta C({\vec{x}})_{\rm stat}^4] - \mathbb{E}[\delta C({\vec{x}})_{\rm stat}^2]^2) \\
&= 3 \sigma_{\textrm{stat}, C}^4 - (\sigma_{\textrm{stat}, C}^2)^2 \\
&=  2 \sigma_{\textrm{stat}, C}^4
\end{align*}

\noindent \textcircled{\tiny{4}}
\begin{align*}
\mathrm{Var}[-2 \delta C({\vec{x}})_{\rm stat} \delta C(\vec{y})_{\rm stat}] &= 4 (\mathbb{E}[\delta C({\vec{x}})_{\rm stat}^2 \delta C(\vec{y})_{\rm stat}^2] \\
&~~~ - \mathbb{E}[\delta C({\vec{x}})_{\rm stat} \delta C(\vec{y})_{\rm stat}]^2) \\
&= 4 (\mathbb{E}[\delta C({\vec{x}})_{\rm stat}^2]^2 - \mathbb{E}[\delta C({\vec{x}})_{\rm stat}]^4) \\
&= 4 \sigma_{\textrm{stat}, C}^4
\end{align*}

\noindent  \textcircled{\tiny{5}} + \textcircled{\tiny{6}}
\begin{align*}
\mathrm{Var}&[2 (C({\vec{x}})-C(\vec{y})) (\delta C({\vec{x}})_{\rm stat} - \delta C(\vec{y})_{\rm stat})] \\
&=  4 \, (\mathbb{E}[(C({\vec{x}})-C(\vec{y}))^2 (\delta C({\vec{x}})_{\rm stat} - \delta C(\vec{y})_{\rm stat})^2] \\
& ~~~ - \mathbb{E}[(C({\vec{x}})-C(\vec{y})) (\delta C({\vec{x}})_{\rm stat} - \delta C(\vec{y})_{\rm stat})]^2) \\
&= 4 \,\mathbb{E}[(C({\vec{x}})-C(\vec{y}))^2 (\delta C({\vec{x}})_{\rm stat}^2 + \delta C(\vec{y})_{\rm stat}^2 \\
& ~~~ - 2 \delta C({\vec{x}})_{\rm stat} \delta C(\vec{y})_{\rm stat})] \\
&= 8 \, \mathbb{E}[(C({\vec{x}})-C(\vec{y}))^2]\mathbb{E}[\delta C({\vec{x}})_{\rm stat}^2] \\
&= 8 \textrm{SF}(s) \sigma_{\textrm{stat}, C}^2
\end{align*}

\paragraph{Covariance term}\mbox{}\\

\noindent The covariance term is only non-zero when the pairs share a common point, as shown in Lemma A, this happens $N_{p}(s) (N_{nei}(s)-1)$ times. In this case we have (assuming ${\vec{x}}$ and $\vec{y}$ permutation of the indices where points are equal)
\begin{align*}
\frac{1}{N_p(s)^2} &\sum_{\substack{({\vec{x}}, \vec{y}), (\vec{x}^{\prime}, \vec{y}^{\prime}) \in \mathbf{\tilde{S}}_s \\ ({\vec{x}}, \vec{y}) \neq (\vec{x}^{\prime}, \vec{y}^{\prime})}} \textrm{Cov} [\zeta^2_{{\vec{x}}, \vec{y}}, \zeta^2_{\vec{x}^{\prime}, \vec{y}^{\prime}}] \\
&~~~ = \frac{2}{N_p(s)^2} \sum_{\substack{({\vec{x}}, \vec{y}), ({\vec{x}},\vec{y}^{\prime}) \in \mathbf{\tilde{S}}_s \\ \vec{y} \neq \vec{y}^{\prime}}} \textrm{Cov} [\zeta^2_{{\vec{x}}, \vec{y}}, \zeta^2_{\vec{x}, \vec{y}^{\prime}}] \\
\end{align*}

\noindent By definition of the covariance, we get 
\begin{align*}
\textrm{Cov} [\zeta^2_{{\vec{x}}, \vec{y}}, \zeta^2_{\vec{x}, \vec{y}^{\prime}}] &= \mathbb{E}[\zeta^2_{{\vec{x}}, \vec{y}} \zeta^2_{\vec{x}, \vec{y}^{\prime}}] - \mathbb{E}[\zeta^2_{{\vec{x}}, \vec{y}}]\mathbb{E}[\zeta^2_{\vec{x}, \vec{y}^{\prime}}]  \\
\end{align*}

\noindent \textbf{Cross-expectation:}
\begin{align*}
\mathbb{E}[\zeta^2_{{\vec{x}}, \vec{y}} \zeta^2_{\vec{x}, \vec{y}^{\prime}}] &= \mathbb{E}[((C({\vec{x}})-C(\vec{y}))^2 + \delta C({\vec{x}})_{\rm stat}^2 + \delta C(\vec{y})_{\rm stat}^2  \\
&~~~ - 2 \delta C({\vec{x}})_{\rm stat} \delta C(\vec{y})_{\rm stat}  \\
&~~~ + 2 (C({\vec{x}})-C(\vec{y})) (\delta C({\vec{x}})_{\rm stat} - \delta C(\vec{y})_{\rm stat})) \\
&~~~ \times ((C({\vec{x}})-C(\vec{y}^{\prime}))^2 + \delta C(
{\vec{x}})_{\rm stat}^2 + \delta C(\vec{y}^{\prime})_{\rm stat}^2 \\
&~~~ - 2 \delta C({\vec{x}})_{\rm stat} \delta C(\vec{y}^{\prime})_{\rm stat} \\
&~~~ + 2 (C({\vec{x}})-C(\vec{y}^{\prime})) (\delta C({\vec{x}})_{\rm stat} - \delta C(\vec{y}^{\prime})_{\rm stat}))]
\end{align*}
Most of the terms that include a linear combination of $\mathbb{E}[\delta C_{\rm stat}]$ or $\mathbb{E}[\delta C_{\rm stat}^3]$ give 0 (as the random field is centred), we remain with
\begin{align*}
\mathbb{E}[\zeta^2_{{\vec{x}}, \vec{y}} \zeta^2_{\vec{x}, \vec{y}^{\prime}}] 
&= \mathbb{E}[\zeta^2_{0, {\vec{x}}, \vec{y}} \zeta^2_{0, \vec{x}, \vec{y}^{\prime}}]  + 4 \sigma_{\textrm{stat}, C}^2 \textrm{SF}(s) \\
&~~~ + 4 \sigma_{\textrm{stat}, C}^2 \mathbb{E}[\zeta_{0,{\vec{x}}, \vec{y}} \zeta_{0,\vec{x}, \vec{y}^{\prime}}]+ 6 \sigma_{\textrm{stat}, C}^4 \\
\end{align*}

\noindent \textbf{Squared-expectation:} When developing the squared expectation, most of the terms are once again 0 given the centred random variable $\delta v_{\rm stat}$. 
\begin{align*}
\mathbb{E}[\zeta^2_{{\vec{x}}, \vec{y}}]\mathbb{E}[\zeta^2_{\vec{x}, \vec{y}^{\prime}}]  
&= \mathbb{E}[\zeta^2_{0, {\vec{x}}, \vec{y}}] \mathbb{E}[\zeta^2_{0, \vec{x}, \vec{y}^{\prime}}] + 4 \sigma_{\textrm{stat}, C}^2 \textrm{SF}(s) + 4 \sigma_{\textrm{stat}, C}^4 \\
\end{align*}

By reassembling all the terms and noticing that $\mathbb{E}[\zeta_{0, {\vec{x}}, \vec{y}}] \mathbb{E}[\zeta_{0, \vec{x}, \vec{y}^{\prime}}] = 0$ the $\mathbb{E}[\zeta_{0,{\vec{x}}, \vec{y}} \zeta_{0,\vec{x}, \vec{y}^{\prime}}]$ term can in fact be expressed as $\textrm{Cov}[\zeta_{0,{\vec{x}}, \vec{y}},\zeta_{0,\vec{x}, \vec{y}^{\prime}}]$. When the terms giving 0 are also included:

\begin{align*}
\frac{2}{N_p(s)^2} &\sum_{\substack{({\vec{x}}, \vec{y}), ({\vec{x}},\vec{y}^{\prime}) \in \mathbf{\tilde{S}}_s \\ \vec{y} \neq \vec{y}^{\prime}}} \textrm{Cov} [\zeta^2_{{\vec{x}}, \vec{y}}, \zeta^2_{\vec{x}, \vec{y}^{\prime}}] \\
&= \underbrace{\frac{1}{N_p(s)^2} \sum_{\substack{({\vec{x}}, \vec{y}), (\vec{x}^{\prime}, \vec{y}^{\prime}) \in \mathbf{\tilde{S}}_s \\ ({\vec{x}}, \vec{y}) \neq (\vec{x}^{\prime}, \vec{y}^{\prime})}} \textrm{Cov} [\zeta^2_{0, {\vec{x}}, \vec{y}}, \zeta^2_{0, \vec{x}^{\prime}, \vec{y}^{\prime}}]}_{(1)} \\
&~~~ + \underbrace{\frac{4 \sigma_{\textrm{stat}, C}^2}{N_p(s)^2} \sum_{\substack{({\vec{x}}, \vec{y}), (\vec{x}^{\prime}, \vec{y}^{\prime}) \in \mathbf{\tilde{S}}_s \\ ({\vec{x}}, \vec{y}) \neq (\vec{x}^{\prime}, \vec{y}^{\prime})}} \textrm{Cov} [\zeta_{0, {\vec{x}}, \vec{y}}, \zeta_{0, \vec{x}^{\prime}, \vec{y}^{\prime}}]}_{(2)} \\
&~~~ + \frac{4(N_{nei}(s)-1)}{N_p(s)} \sigma_{\textrm{stat}, C}^4
\end{align*}

The first term (1) can be assembled with the corresponding variance term \textcircled{\tiny{1}} to obtain $\textrm{Var} (\textrm{SF}(s))$.
\begin{align*}
\frac{1}{N_p(s)^2} &\sum_{({\vec{x}}, \vec{y}) \in \mathbf{\tilde{S}}_s} \mathrm{Var}[\zeta_{0, {\vec{x}}, \vec{y}}^2] \\
&~~~ + \frac{1}{N_p(s)^2} \sum_{\substack{({\vec{x}}, \vec{y}), (\vec{x}^{\prime}, \vec{y}^{\prime}) \in \mathbf{\tilde{S}}_s \\ ({\vec{x}}, \vec{y}) \neq (\vec{x}^{\prime}, \vec{y}^{\prime})}} \textrm{Cov} [\zeta^2_{0, {\vec{x}}, \vec{y}}, \zeta^2_{0, \vec{x}^{\prime}, \vec{y}^{\prime}}] \\
&= \frac{1}{N_p(s)^2} \textrm{Var} [\sum_{({\vec{x}}, \vec{y}) \in \mathbf{\tilde{S}}_s} \zeta^2_{0, {\vec{x}}, \vec{y}}] \\
&= \textrm{Var}[\textrm{SF}(s)]
\end{align*}

The second term (2) can in turn be coupled to a corresponding variance term, by noticing that $\textrm{Var}[\zeta_{0, {\vec{x}}, \vec{y}}] = \mathbb{E}[\zeta_{0, {\vec{x}}, \vec{y}}^2] -  \mathbb{E}[\zeta_{0, {\vec{x}}, \vec{y}}]^2 = \textrm{SF}(s)$. In which case:

\begin{align*}
\textrm{Var}[\textrm{D}(s)] &= \frac{1}{N_p(s)^2}  \sum_{\substack{({\vec{x}}, \vec{y}), (\vec{x}^{\prime}, \vec{y}^{\prime}) \in \mathbf{\tilde{S}}_s \\ ({\vec{x}}, \vec{y}) \neq (\vec{x}^{\prime}, \vec{y}^{\prime})}} \textrm{Cov} [\zeta_{0, {\vec{x}}, \vec{y}}, \zeta_{0, \vec{x}^{\prime}, \vec{y}^{\prime}}] \\
&~~~ +  \frac{1}{N_{p}(s)} \textrm{SF}(s)
\end{align*}
Finally, by subtracting the $\textrm{SF}(s)$ to the left-hand side and by regrouping all the other terms (also from the variance) we obtain that
\begin{align*}
\textrm{Var}[\overline{\textrm{SF}}(s)] &= \textrm{Var}[\textrm{SF}(s)] + 4\textrm{Var}[\textrm{D}(s)]\sigma_{\textrm{stat}, C}^2 +  \\
&~~~  \frac{4}{N_p(s)} \textrm{SF}(s) \sigma_{\textrm{stat}, C}^2 + \frac{4(N_{nei}(s)+1)}{N_{p}(s)} \sigma_{\textrm{stat}, C}^4
\end{align*}

\noindent \textcolor{black}{One should bear in mind that Var[D] and SF are not mutually independent as the former is related to the latter through a SF/$N_p$ term, which adds to the covariance term $\textrm{Cov} [\zeta_{0, {\vec{x}}, \vec{y}}, \zeta_{0, \vec{x}^{\prime}, \vec{y}^{\prime}}]$. Both scale with $\sigma_{\textrm{stat}, C}^2$. We decided to split these terms for two (mathematical) reasons:
\begin{itemize}
\item Uniting the covariance term to a known SF term allows the simplification of the expression into the variance of a single quantity, D. Current work is ongoing to determine analytical expression, similarly to CL19. This would be more complicated than for the covariance term exclusively.
\item Unlike the SF/$N_p$ term, Var[D] does not go to 0 for a large number of pairs. It represents the intrinsic contribution of the cross-product between the statistics and the properties of the field, which bring a systematic error to the estimation, regardless of the number of pairs used to estimate it.
\end{itemize} 
Together, they form a higher order term that provides the interaction between the statistics and the intrinsic properties of the turbulent power spectrum. Due to their different asymptotic behaviour, we chose however to separate them.}

\subsection{Expected variance of $\textrm{D}$}
\label{app:vard}

\noindent The variance of the estimator $\overline{\textrm{D}}$ is defined as 
\begin{align*}
\mathrm{Var}[\overline{\textrm{D}}(s)] &= \mathrm{Var}[\sum_{({\vec{x}}, \vec{y}) \in \mathbf{\tilde{S}}_s} \frac{[(C({\vec{x}})-C(\vec{y}))+\delta C({\vec{x}})_{\rm stat} - \delta C(\vec{y})_{\rm stat}]}{N_p(s)}] \\
&=  \frac{1}{N_p(s)^2} \mathrm{Var}[\sum_{({\vec{x}}, \vec{y}) \in \mathbf{\tilde{S}}_s}  \zeta_{\vec{x},\vec{y}}] \\
&= \frac{1}{N_p(s)^2} ( \sum_{({\vec{x}}, \vec{y}) \in \mathbf{\tilde{S}}_s} \mathrm{Var}[\zeta_{{\vec{x}}, \vec{y}}]+\sum_{\substack{({\vec{x}}, \vec{y}), ({\vec{x}^{\prime}}, {\vec{y}^{\prime}}) \in \mathbf{\tilde{S}}_s \\ ({\vec{x}}, \vec{y}) \neq ({\vec{x}^{\prime}}, {\vec{y}^{\prime}})}} \textrm{Cov} [\zeta_{{\vec{x}}, \vec{y}}, \zeta_{{\vec{x}^{\prime}}, {\vec{y}^{\prime}}}] )\\
\end{align*}

\paragraph{Variance term}\mbox{}\\
\begin{align*}
\frac{1}{N_p(s)^2} \sum_{({\vec{x}}, \vec{y}) \in \mathbf{\tilde{S}}_s} \mathrm{Var}[\zeta_{{\vec{x}}, \vec{y}}] &= \frac{1}{N_p(s)^2} \sum_{({\vec{x}}, \vec{y}) \in \mathbf{\tilde{S}}_s} \mathrm{Var}[ (C({\vec{x}})-C(\vec{y}))  \\
&~~~ + \delta C({\vec{x}})_{\rm stat} - \delta C(\vec{y})_{\rm stat}]
\end{align*}

\noindent \textcircled{\tiny{1}}
\begin{align*}
\frac{1}{N_p(s)^2} \sum_{({\vec{x}}, \vec{y}) \in \mathbf{\tilde{S}}_s} \mathrm{Var}[(C({\vec{x}})-C(\vec{y}))] = \frac{1}{N_p(s)^2} \sum_{({\vec{x}}, \vec{y}) \in \mathbf{\tilde{S}}_s} \mathrm{Var}[\zeta_{0, {\vec{x}}, \vec{y}}]
\end{align*}

\noindent \textcircled{\tiny{2}} / \textcircled{\tiny{3}}
\begin{align*}
\mathrm{Var}[\delta C({\vec{x}})_{\rm stat}] &= ( \mathbb{E}[\delta C({\vec{x}})_{\rm stat}^2] - \mathbb{E}[\delta C({\vec{x}})_{\rm stat}]^2) \\
&= \sigma_{\textrm{stat}, C}^2
\end{align*}

\paragraph{Covariance term}\mbox{}\\

\noindent As in the previous case, we can rewrite the covariance as
\begin{align*}
\frac{1}{N_p(s)^2} &\sum_{\substack{({\vec{x}}, \vec{y}), (\vec{x}^{\prime}, \vec{y}^{\prime}) \in \mathbf{\tilde{S}}_s \\ ({\vec{x}}, \vec{y}) \neq (\vec{x}^{\prime}, \vec{y}^{\prime})}} \textrm{Cov} [\zeta_{{\vec{x}}, \vec{y}}, \zeta_{\vec{x}^{\prime}, \vec{y}^{\prime}}] \\
&~~~ = \frac{2}{N_p(s)^2} \sum_{\substack{({\vec{x}}, \vec{y}), ({\vec{x}},\vec{y}^{\prime}) \in \mathbf{\tilde{S}}_s \\ \vec{y} \neq \vec{y}^{\prime}}} \textrm{Cov} [\zeta_{{\vec{x}}, \vec{y}}, \zeta_{\vec{x}, \vec{y}^{\prime}}] \\
\end{align*}

\noindent By definition
\begin{align*}
\textrm{Cov} [\zeta_{{\vec{x}}, \vec{y}}, \zeta_{\vec{x}, \vec{y}^{\prime}}] &= \mathbb{E}[\zeta_{{\vec{x}}, \vec{y}} \zeta_{\vec{x}, \vec{y}^{\prime}}] - \mathbb{E}[\zeta_{{\vec{x}}, \vec{y}}]\mathbb{E}[\zeta_{\vec{x}, \vec{y}^{\prime}}]  \\
\end{align*}

\noindent \textbf{Cross-expectation:}
\begin{align*}
\mathbb{E}[\zeta_{{\vec{x}}, \vec{y}} \zeta_{\vec{x}, \vec{y}^{\prime}}] &= \mathbb{E}[(C({\vec{x}})-C(\vec{y})(C({\vec{x}})-C(\vec{y}^{\prime}) + \delta C(\vec{y})_{\rm stat}\delta C(\vec{y}^{\prime})_{\rm stat} \\
&~~~ + (C({\vec{x}})-C(\vec{y}))(\delta C({\vec{x}})_{\rm stat}-\delta C(\vec{y}^{\prime})_{\rm stat}) \\
&~~~ + (C({\vec{x}})-C(\vec{y}^{\prime}) (\delta C({\vec{x}})_{\rm stat}-\delta C(\vec{y})_{\rm stat}) \\
&~~~ + \delta C({\vec{x}})_{\rm stat}^2 - \delta C({\vec{x}})_{\rm stat} (\delta C({\vec{y}})_{\rm stat}+\delta C(\vec{y}^{\prime})_{\rm stat}) ]
\end{align*}
By taking only the non-zero terms, we get
\begin{align*}
\mathbb{E}[\zeta_{{\vec{x}}, \vec{y}} \zeta_{\vec{x}, \vec{y}^{\prime}}] 
&= \mathbb{E}[\zeta_{0, {\vec{x}}, \vec{y}} \zeta_{0, \vec{x}, \vec{y}^{\prime}}]  + \sigma_{\textrm{stat}, C}^2
\end{align*}

\noindent \textbf{Squared-expectation:} When developing the squared expectation, most of the terms are once again 0 and we remain with
\begin{align*}
\mathbb{E}[\zeta_{{\vec{x}}, \vec{y}}]\mathbb{E}[\zeta_{\vec{x}, \vec{y}^{\prime}}]  
&= \mathbb{E}[\zeta_{0, {\vec{x}}, \vec{y}}] \mathbb{E}[\zeta_{0, \vec{x}, \vec{y}^{\prime}}] = 0
\end{align*}

\noindent By reassembling all the terms we find:

\begin{align*}
\frac{2}{N_p(s)^2} &\sum_{\substack{({\vec{x}}, \vec{y}), ({\vec{x}},\vec{y}^{\prime}) \in \mathbf{\tilde{S}}_s \\ \vec{y} \neq \vec{y}^{\prime}}} \textrm{Cov} [\zeta_{{\vec{x}}, \vec{y}}, \zeta_{\vec{x}, \vec{y}^{\prime}}] \\
&= \frac{1}{N_p(s)^2} \sum_{\substack{({\vec{x}}, \vec{y}), (\vec{x}^{\prime}, \vec{y}^{\prime}) \in \mathbf{\tilde{S}}_s \\ ({\vec{x}}, \vec{y}) \neq (\vec{x}^{\prime}, \vec{y}^{\prime})}} \textrm{Cov} [\zeta_{0, {\vec{x}}, \vec{y}}, \zeta_{0, \vec{x}^{\prime}, \vec{y}^{\prime}}] \\
&~~~ + \frac{2(N_{nei}(s)-1)}{N_p(s)} \sigma_{\textrm{stat}, C}^4
\end{align*}

\noindent Finally, by including the variance terms, we obtain

\begin{align*}
\textrm{Var}[\overline{\textrm{D}}(s)] &= \textrm{Var}[\textrm{D}(s)] + \frac{2(N_{nei}(s))}{N_{p}(s)} \sigma_{\textrm{stat}, C}^4
\end{align*}




\end{document}